\documentclass[useAMS,usenatbib]{mn2e}
\usepackage{subfigure}
\usepackage{graphicx}

\title[Black Hole Binaries in Star Clusters]{Compact Binaries in Star Clusters I - Black Hole
  Binaries Inside Globular Clusters}
\author[J. M. B. Downing, M. J. Benacquista, M. Giersz, \&
  R. Spurzem]{J. M. B. Downing$^{1,2}$\thanks{E-mail:
  downin@ari.uni-heidelberg.de}, M. J. Benacquista$^{3}$, M. Giersz$^{4}$,
  and R. Spurzem$^{5,6,1}$\\
  $^{1}$Astronomisches Rechen-Institut, Zentrum f\"{u}r Astronomie der
  Universit\"{a}t Heidelberg, Monchhofsra\ss e 12-14,\\ D-69120 Heidelberg,
  Germany\\
  $^{2}$Fellow of the International Max-Planck Research School for Astronomy
  and Cosmic Physics at the University of Heidelberg,\\ Heidelberg, Germany\\
  $^{3}$Center for Gravitational Wave Astronomy, University of Texas at
  Brownsville, Brownsville, TX 78520, USA\\
  $^{4}$Nicolaus Copernicus Astronomical Center, Polish Academy of Sciences,
  ul. Bartycka 18, 00-716 Warsaw, Poland\\
  $^{5}$National Astronomical Observatories, Chinese Academy of Sciences,
  20A Datun Rd., Chaoyang District, 100012, China\\
  $^{6}$Kavli Institute of Astronomy and Astrophysics, Peking University,
  Beijing, China}
\begin{document}
\date{Accepted ... Received ... in original form ...}
\pagerange{\pageref{firstpage}--\pageref{lastpage}} \pubyear{2009}
\maketitle
\label{firstpage}
\begin{abstract}

We study the compact binary population in star clusters, focusing on binaries containing black holes, using a self-consistent Monte Carlo treatment of dynamics and full stellar evolution.  We find that the black holes experience strong mass segregation and become centrally concentrated.  In the core the black holes interact strongly with each other and black hole-black hole binaries are formed very efficiently.  The strong interactions,  however, also destroy or eject the black hole-black hole binaries.  We find no black hole-black hole mergers within our simulations but produce many hard escapers that will merge in the galactic field within a Hubble time.  We also find several highly eccentric black hole-black hole binaries that are potential LISA sources, suggesting that star clusters are interesting targets for space-based detectors.  We conclude that star clusters must be taken into account when predicting compact binary population statistics.

\end{abstract}

\section{Introduction}
\label{intro}

The inspirals and mergers of compact binaries where both members are neutron stars (NS) or black holes (BH) are some of the most promising sources for the current and next generation of ground based gravitational wave (GW) detectors (LIGO, Virgo, GEO 600, TAMA 300).  NS-NS binaries are expected to be the most plentiful merger species in the frequency regime of ground based detectors \citep{AbbottNS,Belczynski07}, however BH-BH binaries are more massive and thus can be detected at larger distances, as far as the Virgo cluster \citep{AbbottBH} for the current generation of gravitational wave detectors and up to cosmological distances for the next generation.  Compact binaries with longer periods and including white dwarfs (WD) may also be detectable in the low-frequency ($5 \times 10^{-5} - 1$ Hz) LISA (Laser Interferometer Space Antenna) frequency band \citep{HBW90,Benacquista01,Nelemans01,BBandB08}.  WD-WD binaries will be the most plentiful stellar mass LISA sources and are expected to produce confusion limited noise (e.g. \citealt{EIS87,HBW90,Nelemans01,Timpano06,Ruiter08}) whereas the less common NS-NS, NS-BH, and BH-BH binaries are potentially resolvable.

In order to make predictions for GW event rates, the population of compact binaries in the universe must be understood.  While the population of NSs in the local universe can be constrained by observations of pulsars (e.g. \citealt{Kalogera01,Lorimer05}), BHs cannot be observed directly and their properties can only be constrained by modelling.  There have been several studies carried out on the compact binary population in the galactic field where stellar and binary evolution proceeds in isolation and can be modelled using simple population synthesis.  In particular \cite{Belczynski07} predict a detection rate for the advanced LIGO detector of $\sim 20$ NS-NS mergers yr$^{-1}$, only $\sim 2$ BH-BH mergers yr$^{-1}$, and $\sim 1$ NS-BH merger yr$^{-1}$.  Thus the detection rate in the galactic field should be dominated by NS-NS mergers.  \cite{BBandB08} have performed similar calculations for the galactic field in the LISA band.  Depending upon the assumptions made about the probability of mergers during common envelope evolution they find $2-6$ resolvable NS-NS binaries and $0-5$ resolvable BH-BH binaries.  This implies that, although rare, both NS-NS and BH-BH binaries may appear as resolved stellar mass sources in the LISA band.  Overall NS-NS binaries will dominate the field detection rate for ground-based detectors while a detection in the LISA band is possible but unlikely.

In star clusters the situation is rather different.  Here interactions between stars and binaries are common (e.g. \citealt{Heggie75}) and can affect the final outcome of binary evolution.  Such interactions can form new binaries from single stars, exchange binary members and field stars and can reduce or increase the period of existing binaries.  In order to maintain energy equipartition, interactions between particles of different masses tend to accelerate the lowest-mass particles to the highest velocities \citep{Spitzer87}.  As a consequence low-mass objects are the most likely to escape during few-body encounters.  Therefore few-body interactions tend to introduce massive objects into binaries.  Massive objects also tend to sink to the center of star clusters where the stellar density is highest (mass segregation, also a consequence of energy equipartion, \citealt{Spitzer87}) and thus massive objects are the most likely to experience dynamical interactions.  Since BHs rapidly become the most massive objects in star clusters due to stellar evolution they will be particularly strongly affected by dense stellar environments and are very likely to be exchanged into binaries.  Therefore star clusters are predicted to form BH-BH binaries rather efficiently \citep{SigPhin93} and may significantly enhance the BH-BH merger rate in the universe.

Several authors have investigated this possibility using various approximations.  \cite{Gultekin04} and \cite{Oleary06} have simulated the formation of intermediate-mass black holes (IMBHs) assuming that the BHs in the cluster are completely mass-segregated and interact only with each other.  In this situation, where the BHs interact very strongly, the authors resolve these encounters with explicit few-body integration.  The interactions can lead to the efficient formation of BH-BH binaries but also tend to destroy or eject those already formed.  These authors conclude that gravitational wave mergers can occur within young star clusters but BH-BH binaries will be destroyed rather efficiently and the population will be depleted within a few Gyrs.  \cite{Oleary06} in particular calculate a star cluster BH-BH detection rate for advanced LIGO of $1-10$ merges yr$^{-1}$, up to 70\% of which actually occur in ejected BH-BH binaries and thus take place in the galactic field.  Neither set of authors include stellar evolution in their simulations.  By contrast \cite{Iva08} and \cite{Sadowski08} have conducted studies of the compact binary population in star clusters using stellar evolution prescriptions and simplified two-zone models of cluster dynamics.  The two-zone models assume that the BHs and BH-BH binaries remain in dynamical equilibrium with the rest of the cluster and do not strongly mass-segregate.  Thus the density of BHs and BH-BH binaries remains low and they do not interact with each other nearly as frequently as in the previous models.  \cite{Sadowski08} in particular find no NS-NS mergers but a much higher rate of BH-BH mergers than \cite{Oleary06}.  They calculate a detection rate of $25-3000$ mergers yr$^{-1}$ for advanced LIGO even though their treatment of the few-body interactions is the same as for \cite{Oleary06}.  This is because although there are fewer interactions that create BH-BH binaries there are also fewer interactions that destroy them.  This highlights the importance assumptions about global cluster dynamics can have on detection rates.  The only study with full treatment of both dynamics and stellar evolution is that of \cite{SPZMcMill00} who conducted small ($N \sim$ a few $10^{3}$) direct N-body simulations and showed that BH-BH binaries are quickly ejected from star clusters due to strong few-body interactions.  This seems to confirm the model of \cite{Oleary06} but the simulations are too small for any strong conclusions to be drawn.  It seems that star clusters can significantly enhance the rate of BH-BH mergers in the universe however by exactly how much depends strongly on the dynamical assumptions made.

All of these simulations resolve the few-body interactions but either use very simplified models for the global cluster dynamics or have values of $N$
too small to represent GCs.  In this paper we use a Monte Carlo code to self consistently model the dynamics of globular clusters over a range of metallicities, binary fractions, and initial concentrations.  We hope to constrain which dynamical assumptions are most likely to produce accurate results for gravitational wave event rate predictions.  We will focus only on the compact binaries that can be found within the cluster during its evolution, leaving a detailed discussion of the population of escapers to a future paper.  In \S~\ref{methods} we briefly describe the Monte Carlo code and some of its features and limitations.  In \S~\ref{IC} we describe our initial models.  In \S~\ref{results} we present the results of our simulations.  In \S~\ref{GWdetect} we present predictions for LISA detections.  We discuss our results in \S~\ref{discussion} and conclude in \S~\ref{conclude}.

\section{Numerical Methods}
\label{methods}

Monte Carlo star cluster simulations use Monte Carlo integration of the theory of two-body relaxation in order to approximate the evolution of GCs.  Assuming a spherically symmetric potential, the orbit of each centre of mass (single star or binary) in the cluster at any instant can be defined by its energy, $E$, and angular momentum vector, $\vec{J}$.  Changes in $E$ and $\vec{J}$ due to the surrounding stars can then be calculated by an appropriate choice of random scattering events from the theory of two-body relaxation.  In this way the dynamical evolution of the star cluster can be simulated self-consistently.  A position for each star can be defined using a time-weighted average over the orbit defined by $E$ and $\vec{J}$ and then the probability of encounters between stars can be calculated.   Unlike other approximate methods for calculating star cluster evolution, each centre of mass is explicitly included in the simulation.  Therefore it is relatively straightforward to incorporate special prescriptions for individual astrophysical events such as few-body interactions (which are not covered by the Monte Carlo approximation) and stellar evolution.

\subsection{The Monte Carlo Code}
\label{code}

We use a H\'enon-type code \citep{Henon71} incorporating the improvements of \cite{Stod82} and \cite{Stod86} for both global and binary dynamics as described by \cite{Giersz98}.  The code includes prescriptions for three-body binary formation and for binary-single and binary-binary encounters.  The probability for three-body binary formation interactions are calculated between all adjacent stars at each timestep according to equations (7) and (8) in \cite{Giersz01}.  New energies, velocities, and orbital parameters are calculated according to \cite{Giersz98}.  Binary-single and binary-binary interaction probabilities are calculated in a similar way to \cite{GandS03}.  The outcome of the binary-binary interactions follow the prescriptions of \cite{Stod86} which are in turn based on the numerical experiments of \cite{Mikkola84}.  Exchange interactions, where one binary member can be exchanged for a field star or the member of another binary, are allowed during binary-single and binary-binary interactions.  The most common exchange outcome is a light star being exchanged for a massive star.  The probability of an exchange interaction for each star is given by equation (17) of \cite{HHandM96}.  Tidal truncation is treated in a simplified way using the approach of \cite{Baumgardt01} \citep{Giersz01,GHandH08}.

Single and binary stellar evolution are simulated using the Single Stellar Evolution (SSE) and Binary Stellar Evolution (BSE) recipes of \cite{HPandT00}
and \cite{HTandP02} \citep{GHandH08}.  These recipes include a full treatment of binary and stellar evolution from the zero age main sequence to the degenerate remnant for a variety of stellar masses and metallicities.  Of particular interest to our work are the natal velocity kicks applied to NSs and
BHs due to asymmetric supernova explosions \citep{LandL94}.  Velocity kicks are applied to all NSs and BHs at birth and are drawn from a Maxwellian velocity
distribution with a dispersion of $\sim 190$ km s$^{-1}$ based on \cite{HandP97}'s proper motion samples of NSs.  The kick velocity of the BHs is then reduced in proportion to the mass of accreted material as described in \cite{Belczynski02}.  The survivability of the binary is calculated as described in \cite{Belczynski06}.  A simplified gravitational wave inspiral timescale in the weak field limit (e.g. \citealt{Peters64}) is also included.

The code has been compared to direct N-body simulations and produces excellent agreement between global dynamical properties for both single \citep{Giersz98} and multi-mass cases (\citealt{Giersz01}, \citealt{Giersz06}).  \cite{GHandH08} have shown the code compares very well with direct N-body simulations when stellar evolution is included.  The code is also able to re-produce several observed physical parameters, such as the surface brightness profiles and luminosity functions, of the observed star clusters M67 \citep{GHandH08}, M4 \citep{HG08} and NGC 6397 \citep{GH09}.

An attractive feature of Monte Carlo simulations is that the computation scales with $\mathcal{O}(N^{1}) - \mathcal{O}(N^{2})$ rather than the $\mathcal{O}(N^{3})$ of direct $N$-body codes.  Furthermore including binaries does not greatly decrease the performance.  This means that a simulation of $N \approx 5 \times 10^{5} - 10^{6}$ bodies with 50\% binaries can be carried out on a single fast processor in the order of hours to days rather than the weeks to months required for direct $N$-body simulations.  At the same time, unlike other approximate methods of calculating star cluster evolution, the Monte
Carlo simulation manages to provide the same star-by-star information produced by direct $N$-body codes.  Thus the Monte Carlo code can be used for large
parameter studies in a short space of time where the details of individual stellar events, such as inspirals and mergers, are of interest.

\subsection{Limitations}
\label{limitations}

The approximations used in the Monte Carlo code introduce limitations that may affect our results.  In particular the BHs may be sufficiently massive
compared to the rest of the system that the become ``Spitzer unstable'' \citep{Spitzer87} and form a decoupled small $N$ subsystem in the cluster core.  This subsystem interacts only with itself and in such a situation if $N$ becomes small enough the distinction between large and small angle scattering breaks down.  The Monte Carlo approximation relies on the fact that these scales may be separated (the cluster can be divided into a near and far zone) and may become unreliable for small Spitzer unstable subsystems.  \cite{Giersz01} and \cite{Giersz06} have shown that the Monte Carlo code accurately re-produces binary burning in the cluster core, indicating that the code can resolve the statistical properties of strong interactions in dense regions well.  Furthermore \cite{HG09} have compared direct $N$-Body and Monte Carlo simulations for the case of the cluster NGC 6397 and have shown good agreement both for the escape rate (driven by ejections from the core) and binary energy generation.  Thus despite its limitations, the Monte Carlo code still seems to be able to produce statistically reliable results for strongly interacting regions.  We will also be able to compare our results to those of \cite{Oleary06}, who make the explicit assumption of a Spitzer unstable BH population, in order to constrain this effect.  Furthermore, if there are a large number of black holes in the cluster any Spitzer unstable subsystem that they may form will be large enough that the Monte Carlo approximation remains valid.

Another issue is our treatment of strong few-body interactions.  Our code currently uses analytic cross-sections calculated in \cite{Heggie75}, \cite{Mikkola84}, \cite{Stod86}, and \cite{HHandM96} for the initialisation and outcome of binary-single and binary-binary interactions and three-body binary formation.  For unequal mass cases these cross-sections are not certain and only allow a limited range of outcomes.  Another problem is that lacking explicit orbital integration, mergers are only possible if the stellar radii overlap at the conclusion of an interaction, ignoring the effect of close approaches during the interaction.  Thus we will probably underestimate the number of compact binary mergers in our simulation.  We note that ideally we would include a few-body integrator in the Monte Carlo code as has been done by \cite{Fregeau07} or in the context of a gas-Monte Carlo hybrid code by \cite{GandS03}.  Such work is planned for the future but will inevitably slow down the code considerably, making large parameter studies more difficult.

\section{Cluster Models}
\label{IC}

\begin{table}
\centering
\caption[InitialCondtions]{The initial conditions for our simulations.  Column 1 gives the model, column 2 the initial binary fraction, column 3 the   metallicity, column 4 the initial ratio of tidal to half-mass radius, column 5 the initial mass, and column 6 the initial half-mass relaxation time.  Both columns 5 and 6 are averaged all ten independent realisations.\label{initcond}}
\scriptsize{
\begin{tabular}[c]{l r r r r r}
\hline
\multicolumn{6}{c}{Initial conditions}\\
\hline
Simulation & $f_{b}$ & $Z$ & $r_{t}/r_{h}$ & $M [{\rm M}_{\odot}]$ & $t_{rh}$
[Myr] \\
\hline
10sol21  & $0.1$ & $0.02$  &  $21$ & $3.61 \times 10^{5}$ & $3.54 \times 10^{3}$ \\
10sol37  & $0.1$ & $0.02$  &  $37$ & $3.63 \times 10^{5}$ & $1.51 \times 10^{3}$ \\
10sol75  & $0.1$ & $0.02$  &  $75$ & $3.62 \times 10^{5}$ & $5.25 \times 10^{2}$ \\
10sol180 & $0.1$ & $0.02$  & $180$ & $3.63 \times 10^{5}$ & $1.41 \times 10^{2}$ \\
50sol21  & $0.5$ & $0.02$  &  $21$ & $5.08 \times 10^{5}$ & $2.99 \times 10^{3}$ \\
50sol37  & $0.5$ & $0.02$  &  $37$ & $5.08 \times 10^{5}$ & $1.28 \times 10^{3}$ \\
50sol75  & $0.5$ & $0.02$  &  $75$ & $5.06 \times 10^{5}$ & $4.44 \times 10^{2}$ \\
50sol180 & $0.5$ & $0.02$  & $180$ & $5.09 \times 10^{5}$ & $1.19 \times 10^{2}$ \\
10low21  & $0.1$ & $0.001$ &  $21$ & $3.60 \times 10^{5}$ & $3.55 \times 10^{3}$ \\
10low37  & $0.1$ & $0.001$ &  $37$ & $3.62 \times 10^{5}$ & $1.51 \times 10^{3}$ \\
10low75  & $0.1$ & $0.001$ &  $75$ & $3.62 \times 10^{5}$ & $5.25 \times 10^{2}$ \\
10low180 & $0.1$ & $0.001$ & $180$ & $3.63 \times 10^{5}$ & $1.41 \times 10^{2}$ \\
50low21  & $0.5$ & $0.001$ &  $21$ & $5.08 \times 10^{5}$ & $2.99 \times 10^{3}$ \\
50low37  & $0.5$ & $0.001$ &  $37$ & $5.07 \times 10^{5}$ & $1.28 \times 10^{3}$ \\
50low75  & $0.5$ & $0.001$ &  $75$ & $5.07 \times 10^{5}$ & $4.44 \times 10^{2}$ \\
50low180 & $0.5$ & $0.001$ & $180$ & $5.07 \times 10^{5}$ & $1.19 \times 10^{2}$ \\
\hline
\end{tabular}
}
\end{table}

We have preformed simulations of star clusters with 16 different sets of initial conditions.  Each simulation has $5.0 \times 10^{5}$ centres of mass(single stars or binaries).  All simulations use a Kroupa initial mass function (IMF) \citep{KroupaIMF}, a broken power-law with a low-mass slope of
$\alpha_{l} = 1.3$, a high-mass slope of $\alpha_{h} = 2.3$ and a break mass of $M_{\rm break} = 0.5 M_{\odot}$.  We follow \cite{Sadowski08} and choose
masses between 0.1 $M_{\odot}$ and 150 $M_{\odot}$.  All simulations are initialised as Plummer models with a tidal cut-off at $r_{tc} = 150$ pc.  According to the classical formula of \cite{Spitzer87}:
\begin{equation}
  \label{rt}
  r_{tc}^{3} = \frac{M_{C}}{2M_{G}}R_{G}^{3}
\end{equation}
where $M_{C}$ is the mass of the cluster, $M_{G}$ is the mass of the galaxy, and $R_{G}$ is the galactocentric radius.  For a galactic mass of $\approx 6 \times 10^{10} M_{\odot}$ and our cluster masses (Table~\ref{initcond}) this yields $R_{G} \sim 9-10$ kpc, a distance just beyond the solar orbit.  Since
we do not include disk shocking in our models, these represent halo clusters.  We choose relatively isolated initial conditions to ensure that the effects we observe are due to internal cluster dynamics and yet can still investigate escapers.  The tidal cut-off is not held constant during the evolution of the cluster but is re-calculated at each timestep according to the current cluster mass.  There is an $N$-dependent parameter, $\alpha$, that describes a
modification of the tidal radius necessary to produce an agreement in escape rate between direct $N$-body models and the Monte Carlo code \citep{Baumgardt01,GHandH08}.  It is set to 1.31 in our simulations which is the value chosen for the M4 models described in \cite{HG08}.  These models have a similar number of particles to the simulations described in this paper and agree well with observations.

We use two different metallicities for our simulations: $Z_{h} = 0.02$ and $Z_{l} = 0.001$.  $Z_{h}$ corresponds roughly to solar metallicity while
$Z_{l}$ corresponds both to the low-metallicity peak of the galactic globular cluster distribution and, for comparison purposes, to the metallicity chosen by \cite{Sadowski08}.  These two metallicities also fall within the high- and low-mass peaks in the bimodal metallicity distribution found for brightest cluster galaxies by \cite{Harris06}.  For the purpose of our study the primary difference between these metallicities is the treatment of stellar mass-loss
in the BSE stellar evolution code.  The mass of single BHs is calculated according to \cite{Belczynski02}.  In these prescriptions mass-loss is suppressed at low metallicity due to less efficient line-driving of stellar winds and this allows the formation of significantly more massive BHs.  These high-mass BHs mass-segregate more swiftly than their low-mass counterparts and will be stronger gravitational wave sources.  The initial distribution of BH masses these prescriptions yield is studied in detail in \cite{Belczynski06}.

\begin{table}
\centering
\caption[Individual BHs]{Number of BHs formed in each model by stellar evolutionary processes.  $N_{sBH}$ is the total number of black holes formed in the cluster, $N_{bBH}$ the total number of black holes formed in binaries.  $N_{BHBH}$ is the total number of black hole-black hole binaries formed by stellar evolutionary processes and $N_{surv}$ is the number of binaries that form a single black hole and survive the formation process.  All primordial BH-BH binaries are disrupted during the second supernovae.  Each quantity is averaged over all ten independent realisations and includes the rms scatter.\label{evBHs}}
\scriptsize{
\begin{tabular}[c]{l r r r r}
\hline
\multicolumn{5}{c}{Individual Black Hole Statistics}\\
\hline
Simulation & $N_{sBH} \pm \sigma$ & $N_{bBH} \pm \sigma$ & $N_{BHBH} \pm
\sigma$ & $N_{surv} \pm \sigma$ \\
\hline
10sol21  & $1103 \pm 18$ &  $197 \pm  8$ &  $2 \pm 1$ & $0 \pm 0$ \\
10sol37  & $1119 \pm 44$ &  $206 \pm 17$ &  $3 \pm 2$ & $0 \pm 0$ \\
10sol75  & $1104 \pm 31$ &  $200 \pm 12$ &  $2 \pm 1$ & $0 \pm 0$ \\
10sol180 & $1129 \pm 22$ &  $190 \pm 18$ &  $3 \pm 1$ & $0 \pm 0$ \\
50sol21  & $1495 \pm 43$ &  $976 \pm 35$ & $13 \pm 5$ & $0 \pm 1$ \\
50sol37  & $1515 \pm 32$ &  $989 \pm 28$ & $12 \pm 4$ & $0 \pm 0$ \\
50sol75  & $1498 \pm 28$ &  $978 \pm 29$ &  $9 \pm 2$ & $0 \pm 1$ \\
50sol180 & $1555 \pm 42$ &  $956 \pm 36$ &  $9 \pm 3$ & $0 \pm 0$ \\
10low21  & $1248 \pm 34$ &  $217 \pm 19$ &  $2 \pm 2$ & $0 \pm 1$ \\
10low37  & $1262 \pm 30$ &  $228 \pm 18$ &  $4 \pm 2$ & $0 \pm 1$ \\
10low75  & $1265 \pm 30$ &  $229 \pm 14$ &  $5 \pm 4$ & $0 \pm 1$ \\
10low180 & $1296 \pm 37$ &  $227 \pm 18$ &  $3 \pm 2$ & $0 \pm 1$ \\
50low21  & $1719 \pm 46$ & $1090 \pm 25$ &  $8 \pm 2$ & $3 \pm 2$ \\
50low37  & $1728 \pm 47$ & $1124 \pm 45$ & $17 \pm 4$ & $3 \pm 2$ \\
50low75  & $1731 \pm 37$ & $1125 \pm 39$ & $16 \pm 3$ & $2 \pm 1$ \\
50low180 & $1769 \pm 31$ & $1077 \pm 33$ &  $9 \pm 3$ & $4 \pm 3$ \\
\hline
\multicolumn{5}{c}{With 50 Realisations} \\
\hline
10sol75  & $1099 \pm 29$ &  $201 \pm 14$ &  $2 \pm 1$ & $0 \pm 0$ \\
50low75  & $1746 \pm 38$ & $1132 \pm 29$ & $18 \pm 4$ & $3 \pm 2$ \\
\hline
\end{tabular}
}
\end{table}
\begin{figure}
\centering
\includegraphics[width=0.5\textwidth]{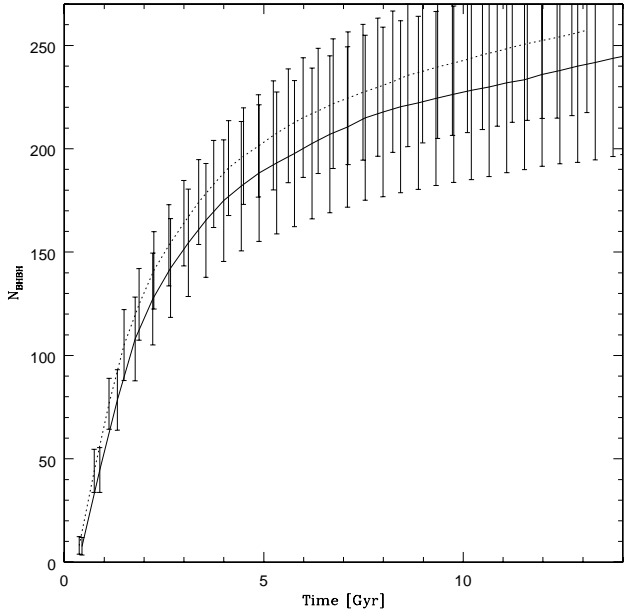}
\caption[Comparison of $\gamma$]{Run 50low75 for two values of $\gamma$:
  $\gamma = 0.02$ (solid) and $\gamma = 0.11$ (dotted).  $\gamma = 0.11$ has a
  slightly shorter relaxation time and produces slightly more BH-BH binaries
  after a Hubble time but both agree to within the rms
  error.\label{fig:gamma}}
\end{figure}

We also use two different binary fractions: $f_{b} = 0.1$ and $f_{b} = 0.5$.  Thus, while they have the same total number of centres of mass, simulations
with $f_{b} = 0.1$ have $5.5 \times 10^{5}$ stars whereas simulations with $f_{b} = 0.5$ have $7.5 \times 10^{5}$ stars.  The simulations with $f_{b} =
0.5$ will be more massive and produce a larger total number of BHs simply because there are more stars present.  $f_{b} = 0.5$ will also produce more binary-single and binary-binary interactions due to the larger number of primordial binaries.  This in turn will increase the probability of exchange interactions and increase the number of BHs introduced into binaries.  The initial binary parameters are produced using the eigenvalue evolution and feeding algorithms of \cite{KroupaBin}.  These prescriptions use a thermal distribution of birth eccentricities ($f(e_{b}) = 2e_{b}$), birth mass ratios ($q_{b}$) drawn at random from the \cite{KroupaIMF} IMF, and a birth period distribution of:
\begin{equation}
\label{eq:PerB}
f(P_{b}) = 2.5\frac{\log{P_{b}} - 1}{45 + (\log{P_{b}} - 1)^{2}}
\end{equation} 
with the limits $\log{P_{b,min}} = 1$ and $\log{P_{b,max}} = 8.43$.  These birth values are then modified according to the eigenvalue and feeding algorithm to simulate the effect of pre-main sequence evolution.  Initial eccentricities are calculated as:
\begin{equation}
\label{eq:eccin}
\ln{e_{in}} = - \rho + \ln{e_{b}}
\end{equation}
where
\begin{equation}
\label{eq:rho}
\rho = \int_{0}^{\Delta t} dt \, \rho^{\prime} = \left( \frac{\lambda
  R_{\odot}}{R_{\rm peri}} \right)^{\chi}
\end{equation}
where $\rho^{\prime -1}$ is the circularisation timescale, $\Delta t \approx 10^{5}$ yr is the pre-main-sequence evolution timescale, $R_{\rm peri}$ is the
pericenter distance of the binary, and $\lambda = 28$ and $\chi = 0.75$ are empirically determined constants.  The initial mass ratio is given by:
\begin{equation}
\label{eq:qin}
q_{in} = q_{b} + (1 - q_{b}) \rho^{\star}
\end{equation}
where
\begin{equation}
\label{eq:rhostar}
\rho^{\star} = \left\{ \begin{array}{ll}
    \rho & \rho \le 1 \\
    1 & \rho > 1
\end{array} \right.
\end{equation}
where the mass of the secondary is modified according to $m_{2,in} = q_{in}m_{2,b}$ and the mass of the primary is unchanged $m_{1,in} = m_{1,b}$.  Finally the period is given by:
\begin{equation}
\label{eq:Pin}
P_{in} = P_{b} \left( \frac{m_{t,b}}{m_{t,in}} \right)^{1/2} \left(
\frac{1-e_{b}}{1-e_{in}} \right)^{3/2}
\end{equation}
where $m_{t,b}$ and $m_{t,in}$ are the total masses before and after the application of Equation~\ref{eq:qin}.  The main effect of the eigenvalue and feeding evolution is to depopulate the short-period, high-eccentricity area of the period-eccentricity diagram as observed in galactic binaries.  The initial period distribution and the effect of the eigenvalue feeding can be seen in figures 1 and 2 of \cite{KroupaBin}.  The effect of a dense stellar evironment on generic binary populations has been investigated in several sources such as \cite{Heggie75} and \cite{OCV}.  The \cite{KroupaBin} prescriptions also provide a good match to the binary parameters observed in the galactic field.

We have performed simulations with four initial concentrations that we control using the ratio of the initial tidal radius to the initial half-mass ($r_{h}$).  We use initial ratios of $r_{t}/r_{h} = 21$, $37$, $75$, and $180$, corresponding to initial number densities within $r_{h}$ of $\sim 10^{2}$, $10^{3}$, $10^{4}$, and $10^{5}$ respectively.  The initial concentration primarily affects the half-mass relaxation time ($t_{rh}$), defined by \cite{Spitzer87} as:
\begin{equation}
  \label{trh}
  t_{rh} = 0.138\frac{N^{1/2}r_{h}^{3/2}}{\langle m \rangle^{1/2}G^{1/2} \ln
    \gamma N}
\end{equation}
Where $N$ is the total number of centres of mass in the system, $\langle m \rangle$ is the average mass, $\ln{\gamma}$ is the Coulomb logarithm, and
$\gamma$ is an empirically fitted parameter.  The value of $\gamma$ is different for single and multimass systems and has been debated in the literature.  For equal mass systems $\gamma = 0.11$ seems to be favoured \citep{GandH94} whereas for multimass systems $\gamma = 0.02$ gives better results \citep{GHandH08}.  We choose $\gamma = 0.02$ for our simulations but have re-run one set, 50low75, with $\gamma = 0.11$ for comparison purposes.  The results will be discussed in \S~\ref{results}.  The simulations all have a fixed initial tidal cut-off and thus the more concentrated simulation have smaller values of $r_{h}$.  Thus, according to Equation~\ref{trh}, they will also have shorter values of $t_{rh}$ and will evolve, dynamically speaking, more quickly than their less concentrated counterparts.  Whatever effect dynamics have on the production of BH-BH binaries should be accelerated in these systems.

Since star cluster dynamics are stochastic and chaotic, large fluctuations can occur between different realisations of the same simulation (e.g. \citealt{GHandH08,HG08,GH09}).  For this reason we perform ten independent realisations of each combination of initial conditions differing only by the initial random seed.  Thus we have a total of 160 simulations to analyse.  To ensure that ten simulations is enough for convergence we have extended two of the simulations, 10sol75 and 50low75, to 50 simulations and will compare the number of BH-BH binaries produced in \S~\ref{results}.  Table~\ref{initcond} gives the initial parameters of our 16 different sets of simulations, averaged over all ten realisations.  Each simulation is run on a single processor at the HLRS supercomputer in Stuttgart.  The shortest simulations (10sol21) take $\sim 4$ h to complete and the longest (50low180) take $\sim 12-16$ h.

\section{Results}
\label{results}

\begin{table}
\centering
\caption[BH-BH binaries]{The cumulative number of BH-BH binaries after $3$, $9$, and $25$ $t_{rh}$, and also after $1 T_{H}$.  Each column is averaged over all ten independent realisations and includes the rms scatter.  A dash in a column indicates that the cluster did not reach that number of half-mass relaxation times within one Hubble time.\label{numBHBH}}
\scriptsize{
\begin{tabular}[c]{l r r r r}
\hline
\multicolumn{5}{c}{BH-BH Binaries After $x t_{rh}$}\\
\hline
Simulation & $t = 3 t_{rh}$ & $t = 9 t_{rh}$ & $t = 25
t_{rh}$ & $t = 14$ Gyr \\
\hline
10sol21  &   $1 \pm  1$ &        -     &        -     &   $1 \pm   1$ \\
10sol37  &   $1 \pm  1$ &  $14 \pm 11$ &        -     &  $14 \pm  11$ \\
10sol75  &   $0 \pm  1$ &   $8 \pm  6$ &  $49 \pm 19$ &  $52 \pm  19$ \\
10sol180 &   $0 \pm  0$ &  $12 \pm  6$ &  $54 \pm 21$ & $123 \pm  27$ \\
50sol21  &   $1 \pm  1$ &        -     &        -     &   $3 \pm   2$ \\
50sol37  &   $3 \pm  2$ &  $36 \pm 10$ &        -     &  $50 \pm  11$ \\
50sol75  &   $1 \pm  1$ &  $26 \pm  8$ & $115 \pm 24$ & $147 \pm  28$ \\
50sol180 &   $0 \pm  1$ &  $11 \pm  5$ & $111 \pm 23$ & $354 \pm  33$ \\
10low21  &  $22 \pm 10$ &        -     &        -     &  $27 \pm  10$ \\
10low37  &  $23 \pm  4$ &  $44 \pm  6$ &        -     &  $44 \pm   6$ \\
10low75  &  $18 \pm 10$ &  $32 \pm 13$ &  $54 \pm  6$ &  $54 \pm  16$ \\
10low180 &  $26 \pm  8$ &  $51 \pm  8$ &  $79 \pm 20$ & $112 \pm  24$ \\
50low21  & $104 \pm 16$ &        -     &        -     & $127 \pm  16$ \\
50low37  &  $93 \pm 22$ & $175 \pm 29$ &        -     & $184 \pm  29$ \\
50low75  &  $64 \pm 10$ & $155 \pm 22$ & $173 \pm 22$ & $202 \pm  38$ \\
50low180 & $103 \pm 19$ & $205 \pm 35$ & $294 \pm 50$ & $453 \pm 109$ \\
\hline
\multicolumn{5}{c}{With 50 Realisations}\\
\hline
10sol75  &   $0 \pm  1$ &   $7 \pm  4$ &  $38 \pm 20$ &  $41 \pm  21$ \\
50low75  &  $78 \pm 15$ & $175 \pm 29$ & $232 \pm 43$ & $245 \pm  48$ \\
\hline
\end{tabular}
}
\end{table}

We find no more than one or two NS-NS or NS-BH binaries over the course of an entire Hubble time in any of our simulations and no NS-NS or NS-BH mergers.  For this reason we only present results for the BH-BH binary population.  The lack of NS-NS and NS-BH binaries is due to the fact that few primordial binaries survive to a phase where they would contain an NS or BH since most merge during mass transfer and those that do are disrupted at the second supernova.  As we will show, our BH-BH binaries are not primordial but rather form dynamically, a process that occurs most efficiently for massive objects.  Since the BHs are, for the most part, significantly more massive than the NSs in our simulations, they will be preferentially exchanged into binaries until they are depleted.  None of our simulations are completely depleted of BHs after one Hubble time and thus the NSs have little opportunity to take part in dynamical compact binary formation.  We do not analyse the (large) WD-WD population but save these results for a future paper.

\begin{figure*}
\centering
\includegraphics[width=\textwidth]{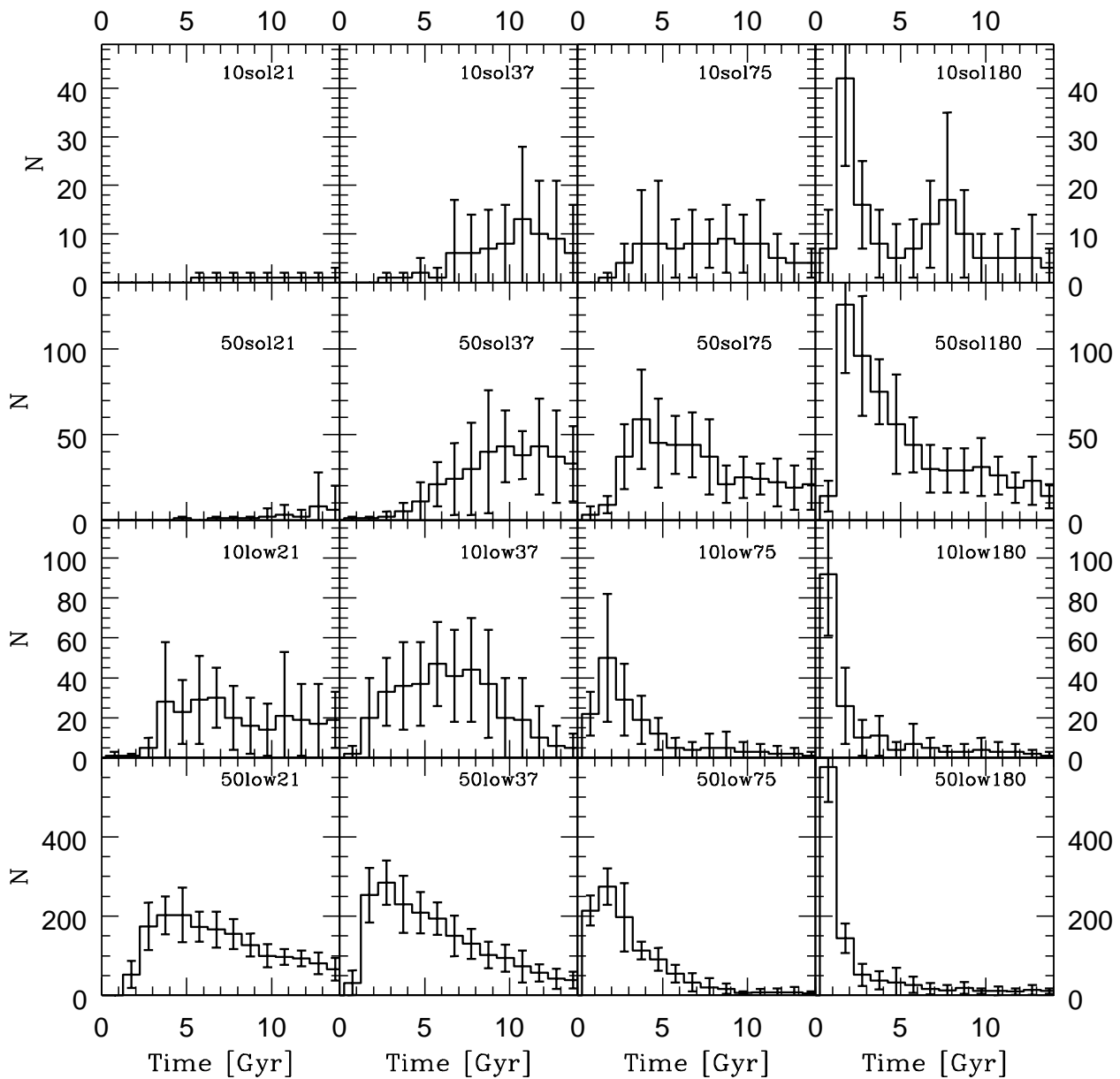}
\caption[BH-BH rates]{The number of BH-BH binaries per Gyr for each of the 16 sets of initial conditions.  Each time bin is averaged over all ten independent realisations and the error bars represent the rms scatter.\label{BHBHrates}}
\end{figure*}

\begin{figure*}
\centering
\includegraphics[width=\textwidth]{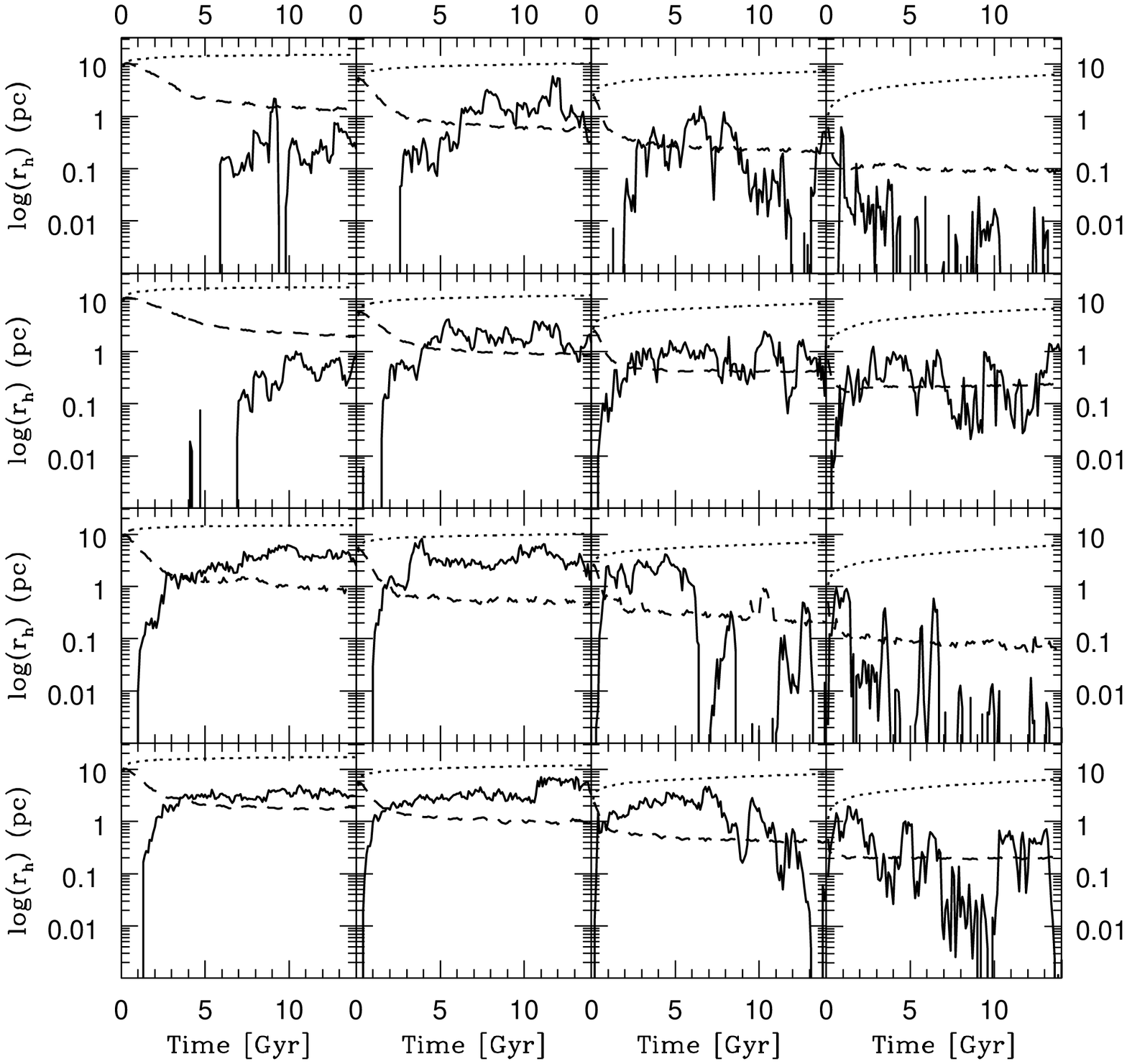}
\caption[$r_{h}$ of BH system]{The half-mass radii for the BH system.  Top to bottom: $Z = 0.02$ and $f_{b} = 0.1$, $Z = 0.02$ and $f_{b} = 0.5$, $Z =
  0.001$ and $f_{b} = 0.1$, and $Z = 0.001$ and $f_{b} = 0.5$.  Left to right: $r_{t}/r_{h} = 21$, $37$, $75$ and $180$.  Shown are cluster $r_{h}$ (dotted), $r_{h}$ of all BHs (dashed), and $r_{h}$ of all BH-BH binaries (solid).  The radial profiles have been boxcar smoothed with a box 100 Myr box and each box has been averaged across all ten independent realisations.\label{radii}}
\end{figure*}

In Table~\ref{evBHs} we show the total number of BHs formed both in isolation and in binaries in each simulation averaged over all ten independent realisations.  For each simulation the rms scatter across the independent realisations is small and is merely a result of random sampling of the IMF.  The number of BHs formed is not a function of concentration in any range of $Z$ or $f_{b}$.  This is not surprising since BHs are produced primarily by
individual stellar evolutionary processes.  It is possible that extra BHs could be produced by stellar collisions in which two stars below the critical mass to produce a BH merge to form a single star above the critical mass.  Collisions would be expected to be more frequent in dense clusters but this effect does not seem to produce a significant enhancement in the number of BHs in our simulations.  The number of BHs formed depends on $f_{b}$ because a higher binary fraction corresponds to a larger number of stars, and on $Z$ because mass-loss is less efficient at low metallicity and stars with a lower zero age main sequence mass can become BHs.  Proportionately more BHs are formed in binaries at $f_{b} = 0.5$ than at $f_{b} = 0.1$ but this is simply a consequence of the larger fraction of stars found in binaries at high $f_{b}$.  It is apparent from column 5 of table~\ref{evBHs} that very few of the binaries that form a single BH survive is formation; most either merge or are disrupted at the supernovae.  Furthermore, very few binaries where both members are black holes form directly from primordial binaries and of those $all$ are disrupted during the formation of the second BH.  Therefore all BH-BH binaries produced by our simulations must be formed by dynamical means.  There is no difference in either the mean number or the rms scatter in BHs produced in either of the simulations for which we have performed an extended number of realisations.  We therefore conclude that ten simulations are sufficient to produce accurate statistics on the BH population produced by stellar evolution.

\begin{figure*}
\centering
\includegraphics[width=\textwidth]{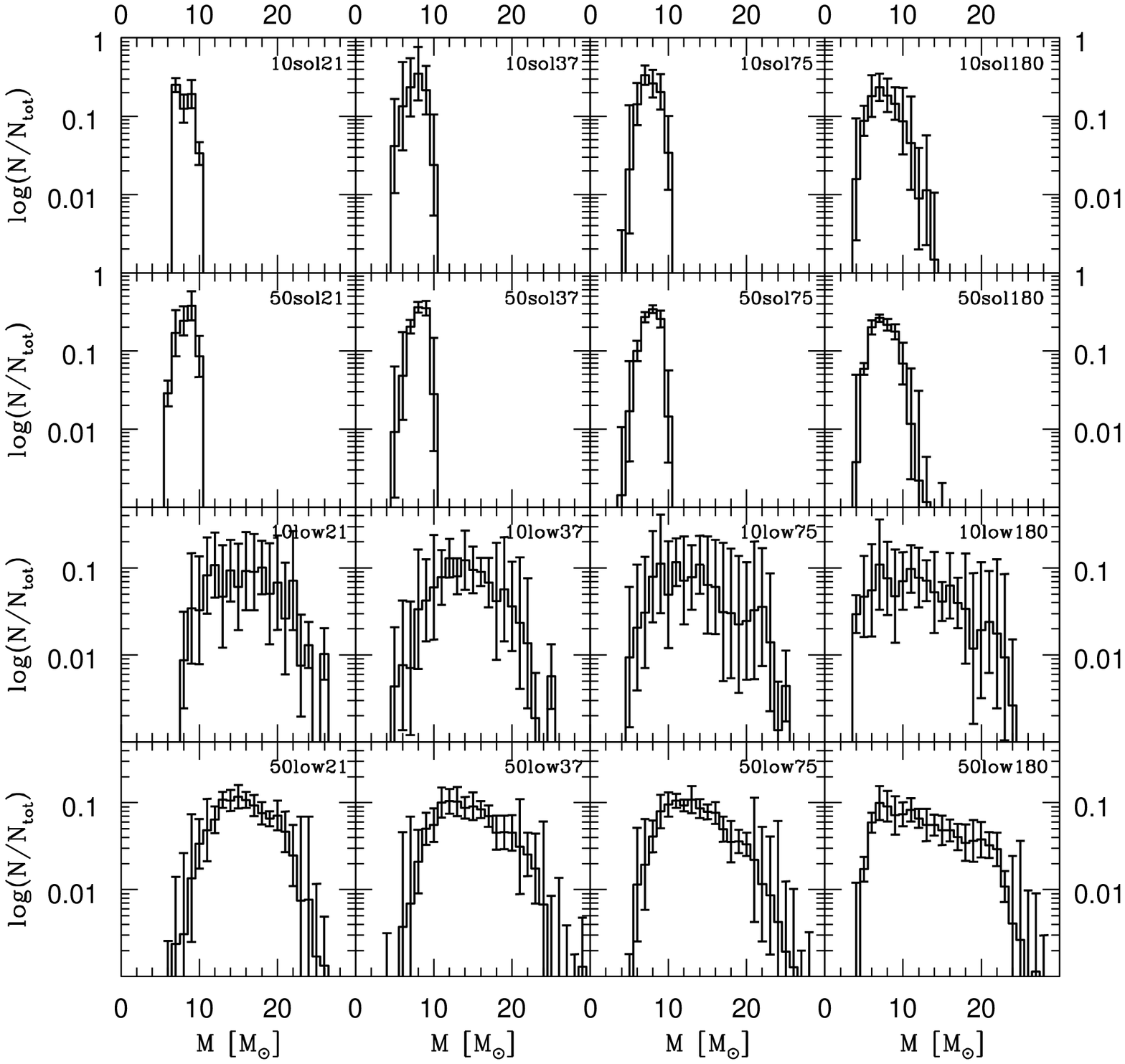}
\caption[Chirp Mass]{The cumulative chirp mass distribution of BH-BH binaries up to $1 T_{H}$ binned in $1 M_{\odot}$ bins.  Each bin in each panel is   averaged over all ten independent realisations and the error bars give the rms scatter.\label{chirp}}
\end{figure*}
\begin{figure*}
\centering
\includegraphics[width=\textwidth]{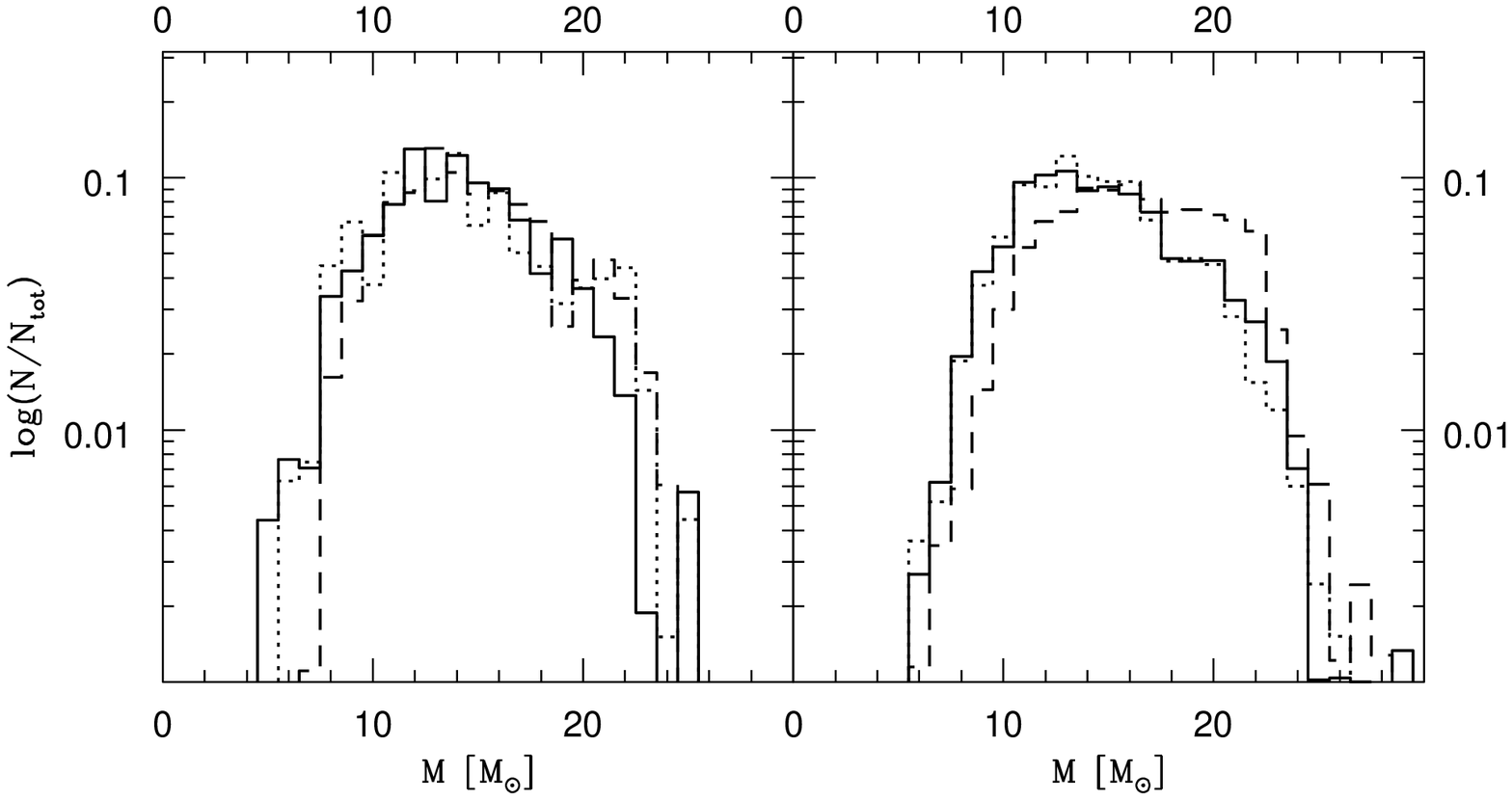}
\caption[Chirp Masses at $9 t_{rh}$]{The cumulative chirp mass distribution of BH-BH binaries up to $9 t_{rh}$ for  six simulations binned in $1 M_{\odot}$ bins.  Both plots are for $Z = 0.001$ with the left plot having $f_{b} = 0.1$ and the right plot having $f_{b} = 0.5$.  Concentrations are $r_{t}/r_{h} = 37$ (solid), $r_{t}/r_{h} = 75$ (dotted), and $r_{t}/r_{h} = 180$ (dashed).  $9 t_{rh}$ is the dynamical age of the simulation with $r_{t}/r_{h} = 37$ after one Hubble time.  The simulations with $r_{t}/r_{h} = 21$ are not shown since they do not reach $9 t_{rh}$ within one Hubble time and have too few BH-BHs at $3 t_{rh}$ for interesting statistics.  Each bin is averaged over all ten independent realisations.\label{chirp9}}
\end{figure*}

In Table~\ref{numBHBH} we present the cumulative number of BH-BH binaries that have existed in each simulation up to the dynamical time shown and after one Hubble time ($T_{H} = 14$ Gyr).  The ages $3 t_{rh}$, $9 t_{rh}$, and $25 t_{rh}$ correspond respectively to the dynamical age of the clusters with $r_{t}/r_{h} = 21$, $37$, and $75$ at $\approx 1 T_{H}$.  The simulations with $r_{t}/r_{h} = 180$ have a dynamical age of $\sim 115-120 t_{rh}$ at $1 T_{H}$.  A new BH-BH binary is counted every time a binary where both members are BHs forms.  Both a binary with one BH and one main sequence (MS) star where the MS star is exchanged for a BH and a binary where both members are BHs and one of the BHs is exchanged for a new BH are counted as new BH-BH binaries.  The rms scatter in the number of BH-BH binaries produced by different realisations of the same simulation in Table~\ref{numBHBH} is significantly larger than for Table~\ref{evBHs}.  This is because, unlike stellar evolution, dynamical binary formation is a stochastic process, strongly dependent on chance encounters that are a function of the detailed dynamics of the specific system, and, since all BH-BH binaries are formed dynamically, large system-to-system variations are expected.  Again we find that for the two simulations where we performed additional independent realisations, the additional realisations provide no difference in the size of the rms scatter and the average number of BH-BH binaries at each relaxation time is the same to within the rms scatter.  Therefore we conclude that ten realisations is also enough to constrain the statistics of the dynamical processes and we retain this number for the rest of our analysis.

There are some clear trends in Table~\ref{numBHBH}.  After $1 T_{H}$ the number of BH-BH binaries increases with $f_{b}$ and initial concentration and decreases with $Z$.  The reasons for the $f_{b}$ correlation are clear: a larger number of both BHs and hard binaries for the BHs to be exchanged into.
The correlation with initial concentration is related to the relative dynamical ages of the clusters after $1 T_{H}$.  For the same $f_{b}$ and $Z$ the simulations have roughly the same number of BH-BHs when compared at the same dynamical age.  By $1 T_{H}$, however, clusters with a higher initial concentration are dynamically older and have had more opportunity to produce BH-BH binaries than their dynamically younger counterparts.  The correlation with $Z$ is due to the higher mass BHs present at low metallicity.  Since the mass-segregation timescale, $t_{eq}$, (for a two-component system) scales as \citep{Spitzer87}:
\begin{equation}
  \label{teq}
  t_{eq} \propto t_{rh}\frac{m_{1}}{m_{2}}
\end{equation}
where $m_{2} > m_{1}$ \citep{Watters00,Khalisi07}, we expect more massive BHs to mass segregate more rapidly than less massive ones.  This accelerates the process of BH-BH binary formation since the core is the densest region and is where most dynamical binary formation will take place.  The more massive BHs are also more likely to be retained by the cluster since they will need larger kicks upon formation in order to reach escape velocity.

We have also considered a different value of $\gamma$ and we present these results in figure~\ref{fig:gamma}.  According to equation~\ref{trh}, the effect of increasing $\gamma$ is to reduce $t_{\rm rh}$ by the logarithm of the same factor.  Thus the simulations with $\gamma = 0.11$ are slightly more dynamically evolved after $1 T_{H}$ that the simulations with $\gamma = 0.02$.  Consequently these simulations produce a few more BH-BH binaries.  The increase, however, is small and the average values for both sets of simulations fall within each others rms scatter.  We conclude that any reasonable changes in $\gamma$ will not affect our results in any important way.

Figure~\ref{BHBHrates} shows the number of BH-BH binaries in each simulation per Gyr.  The trends noted in Table~\ref{numBHBH} are apparent, particularly those associated with the initial concentration.  Simulations with higher initial concentration have a peak number of BH-BH binaries per unit time much earlier than those with lower concentration.  Those with lower concentration sustain more constant but lower BH-BH populations over longer spans of physical time.  It is also clear that the more metal-poor simulations evolve more quickly and produce BH-BH binaries earlier than their metal-rich counterparts.  This again is a consequence of faster mass segregation. Finally, $f_{b}$ does not affect the time of peak BH-BH binary number but simulations with $f_{b} = 0.5$ are able to sustain production of BH-BH binaries longer because the supply of BHs and hard binaries to exchange them into is larger in these clusters.

\begin{figure*}
\centering
\includegraphics[width=\textwidth]{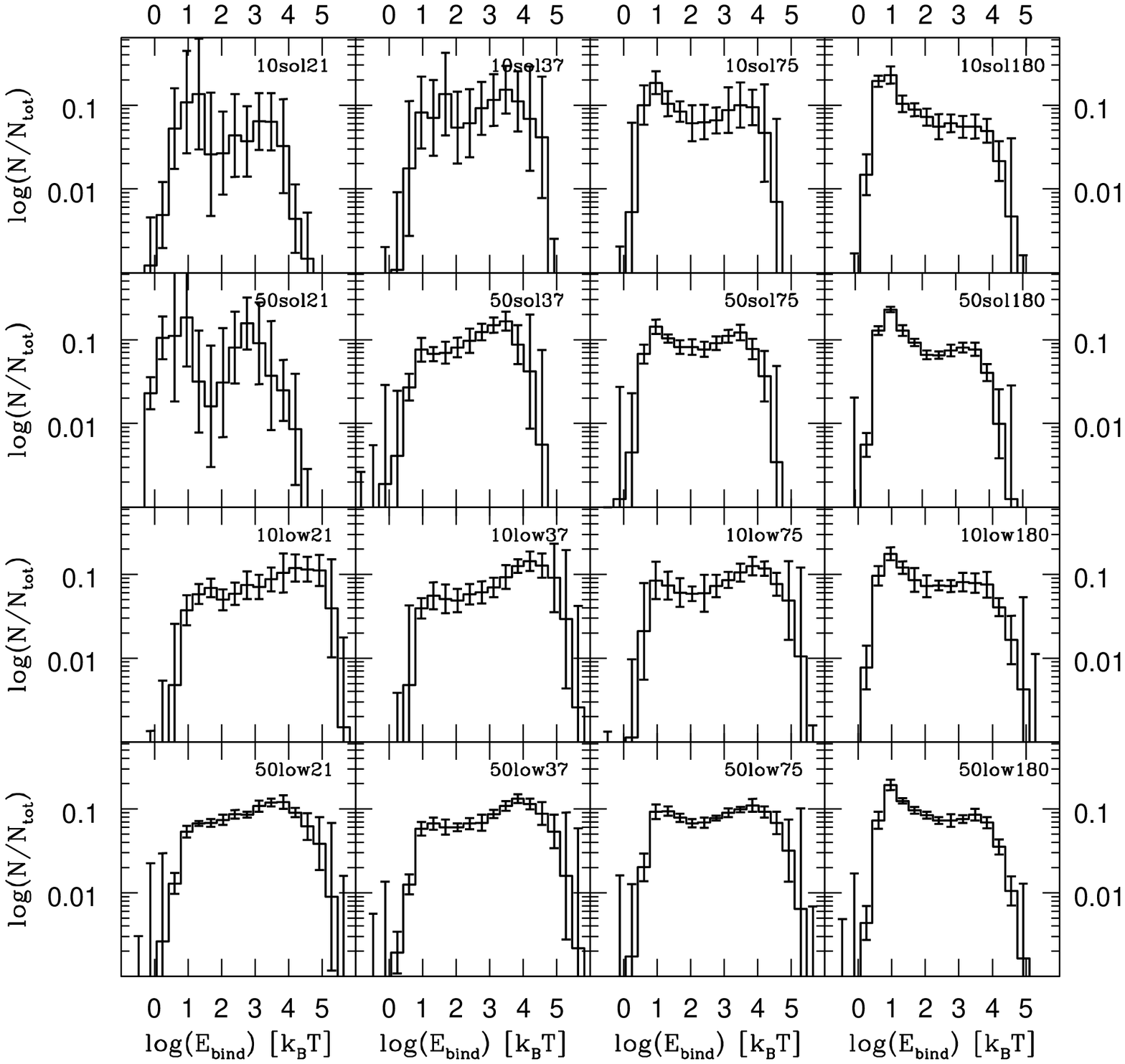}
\caption[Binding Energy of BH-BH Binaries]{Cumulative binding energy distribution of BH-BH binaries up to $1 T_{H}$ binned uniformly in log space.  Each bin is averaged over all ten independent realisations and the error bars give the rms scatter.  Energies are in units of the thermal energy of the core of the cluster.\label{Ebind}}
\end{figure*}

In Figure~\ref{radii} we investigate the spatial distribution of the BHs and BH-BH binaries in each simulation by comparing the half-mass radius of each species to the half-mass radius of the entire cluster.  In all cases single BHs are centrally concentrated compared to the rest of the stars.  This is simply a consequence of mass segregation.  The varying mass segregation timescales can be seen in the length of time taken for the half-mass radii of the BH populations to contract to an equilibrium state caused by binary burning.  The half-mass radii of the BH-BH populations must be interpreted more carefully.  Since BH-BH binaries are formed dynamically in cluster cores, the BH-BH binary populations is initially very centrally concentrated.  As the population evolves, however, the BH-BH binaries interact strongly and can eject each other from the core region.  As is clear from Figure~\ref{BHBHrates} there are often very few BH-BH binaries in the cluster, even over a span of 1 Gyr, and thus the half-mass radius of the BH-BH population in Figure~\ref{radii} is often based on quite a small number of objects.  Thus the location of a single massive BH-BH binary can dominate the determination of the half-mass radius of the BH-BH binary population.  Overall it is clear that BH-BH binaries form in the cluster core and tend to be centrally concentrated but they are not necessarily more concentrated than the single BH population.  Individual BH-BH binaries may also exit in the outskirts of the cluster for a time if they are ejected from the core by dynamical interactions and before they have a chance to sink back to the center due to dynamical friction.

It is also worth noting that the BH-BH binaries will not be strongly affected by interactions with anything other than BHs.  The comparative masses mean that other stars do not have sufficient energy to disrupt or scatter the BH-BH binaries.  Conversely the BH-BH binaries will have a very strong effect on the other stars they encounter and will be a the major energy source in the cluster core.  Thus BH-BH binaries are only affected by each other but have a major influence on core dynamics.

\begin{figure*}
\centering
\includegraphics[width=\textwidth]{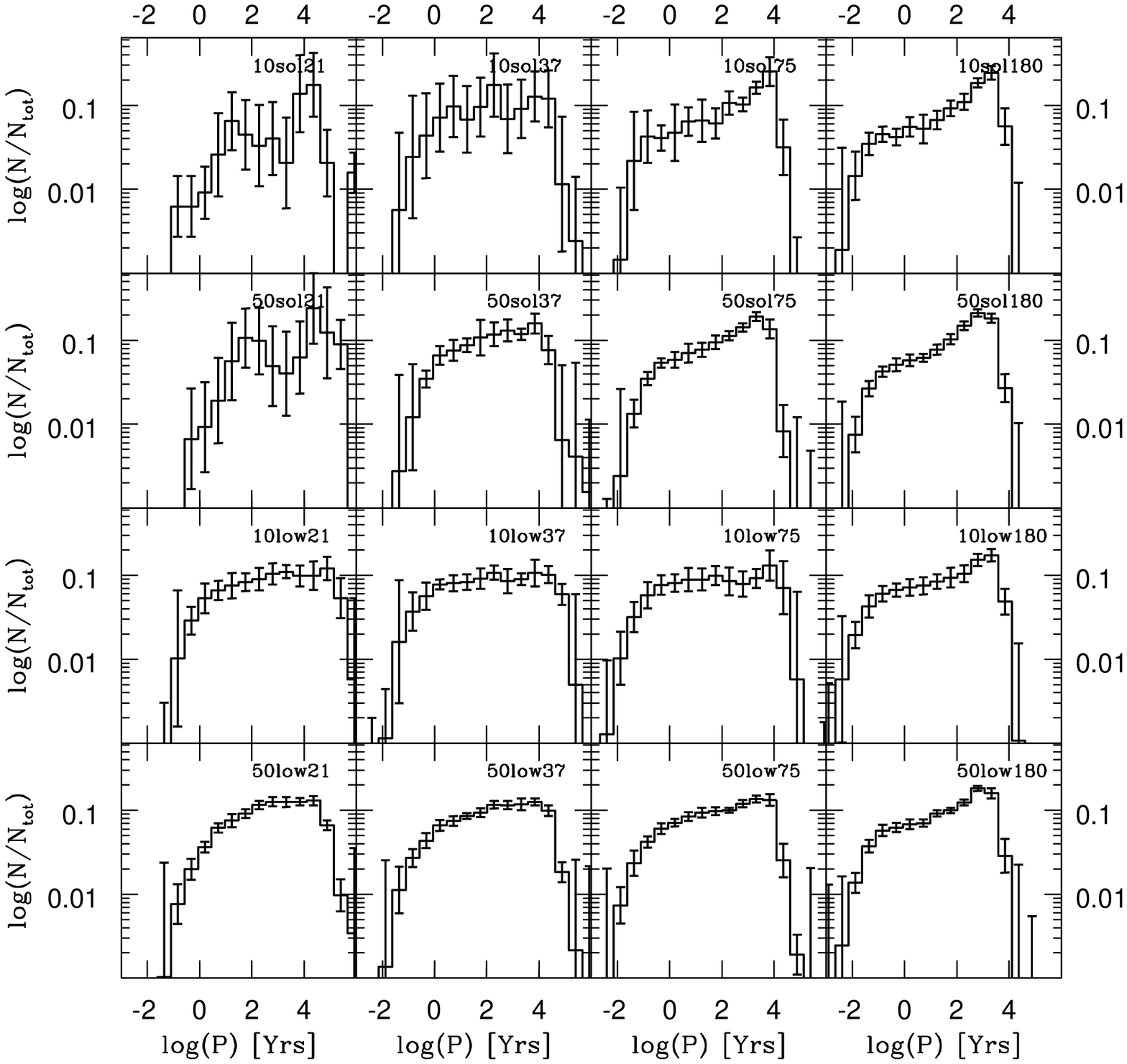}
\caption[Periods of BH-BH Binaries]{Cumulative period distribution of BH-BH binaries up to $1 T_{H}$ binned uniformly in log space.   Each bin is averaged over all ten independent realisations and the error bars give the rms scatter.\label{periods}}
\end{figure*}

The cumulative chirp mass ($M_{\rm chirp} =
(M_{1}M_{2})^{3/5}/(M_{1}+M_{2})^{1/5}$) distribution for all BH-BH binaries that have existed in the simulations up to $1 T_{H}$ are given in Figure~\ref{chirp}.  We choose to display $M_{\rm chirp}$ rather than the total mass because the amplitude of gravitational radiation, $h_{0}$, scales as
\begin{equation}
\label{eq:h0}
h_{0} \propto \frac{G^{5/3}\omega^{2/3}M_{\rm chirp}^{5/3}}{rc^{4}}
\end{equation}
\citep{Pierro01}.  Thus it is $M_{\rm chirp}$ and not the total mass of the binary that is significant for gravitational wave detection.  For the high metallicity simulations the distribution of chirp masses is narrow with a peak around $8-10 M_{\odot}$.  The distribution is not affected by any of the other initial conditions.  The low-metallicity distribution is broader and fairly flat between $10-20 M_{\odot}$.  This is a direct consequence of the more massive BHs generated at lower metallicity.  Here the distribution is weakly affected by the initial concentration with $M_{\rm chirp}$ peaking at lower masses for the more concentrated simulations.  This is a result of the relative dynamical ages of the simulations as we demonstrate in Figure~\ref{chirp9}, the distribution of $M_{\rm chirp}$ after $9 t_{rh}$.  At the same dynamical age the mass distribution is unaffected by the concentration. Recalling Equation~\ref{teq}, the more massive BHs will mass segregate before the less massive BHs and thus interact and be disrupted or ejected earlier.  Only then will low-mass BHs participate in BH-BH binary formation.  Since the more concentrated clusters are dynamically older, they have had more time to experience this effect, deplete their high-mass BHs, and have a lower-mass BH-BH binary population.  The high-metallicity clusters do not have a broad enough distribution in mass for this effect to be important.  Perhaps the most interesting result is that after $1 T_{H}$ the $M_{\rm chirp}$ distributions are systematically different between clusters with different metallicities and concentrations.  Building an $M_{\rm chirp}$ distribution from gravitational wave observations can yield information on the physical and dynamical state of GCs.

\begin{figure*}
\centering
\includegraphics[width=\textwidth]{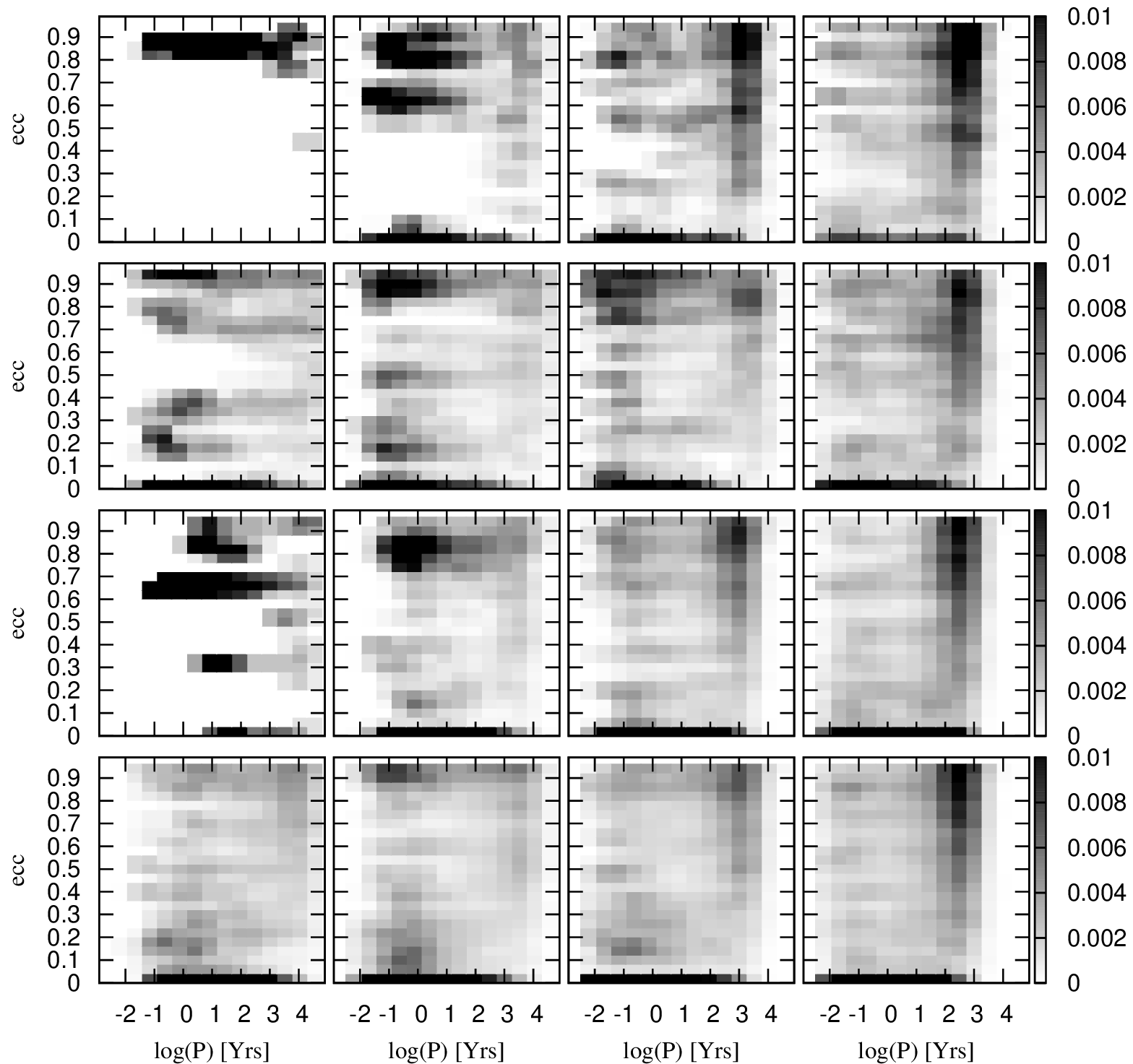}
\caption[Eccentricity vs. Period for BH-BH Binaries]{The eccentricity as a function of period for all BH-BH binaries at all times in all simulations.  From top to bottom $f_{b} = 0.1$ and $Z = 0.02$, $f_{b} = 0.5$ and $Z = 0.02$, $f_{b} = 0.1$ and $Z = 0.001$, and $f_{b} = 0.5$ and $Z = 0.001$.  Form left to right $r_{t}/r_{h} = 21$, $37$, $75$ and $180$.\label{PvsEcc}}
\end{figure*}

The cumulative distribution of BH-BH binary binding energy ($E_{\rm bind}$) up to $1 T_{H}$ is given in Figure~\ref{Ebind}.  The energy is given in units of the thermal energy of the core of the cluster, 
\begin{equation}
  \label{eq:kTcore}
  k_{B}T = (m_{\rm core}/2N_{\rm core})\sigma_{\rm core}^{2}
\end{equation}
where $m_{\rm core}$ is the total core mass, $N_{\rm core}$ the number of stars in the core and $\sigma_{\rm core}$ the velocity dispersion in the core.  The core quantities are chosen because the BH-BH binaries are formed and interact in the core, making this region the most relevant for the dynamics.  All BH-BH binaries are hard.  This is to be expected because soft binaries would be destroyed by the interactions necessary to introduce a BH into them.  There is little variation with the cluster parameters because $k_{B}T$, the normalisation factor for $E_{\rm bind}$, scales with core mass and density.  The only exception to this appears to be clusters with $r_{t}/r_{h} = 180$ where there are an excess of soft binaries.  The combination of the larger interaction cross-section for binaries with larger semi-major axes and the larger interaction rate in the cores of very dense clusters may lead to more BHs being exchanged into softer binaries in the high-density simulations.  It could also be that the soft binaries in the more concentrated simulations still have a larger binding energy in physical units than those in the less concentrated simulations and have a slightly better chance of survival.

\begin{figure*}
\centering
\includegraphics[width=\textwidth]{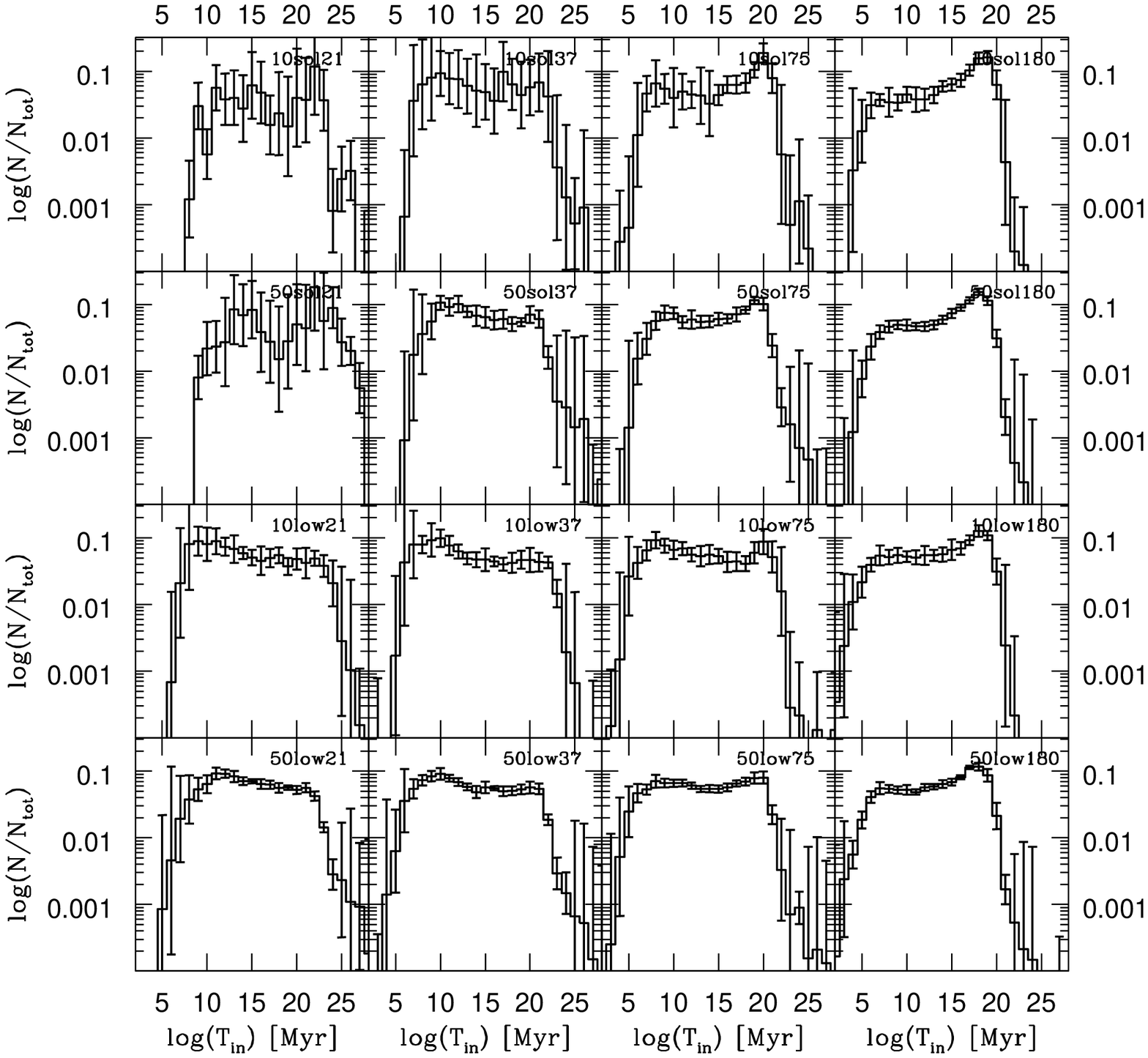}
\caption[Gravitational Wave Inspiral Timescales]{Cumulative distribution of BH-BH binary gravitational wave inspiral timescale up to $1 T_{H}$ binned   uniformly in log space.  Each bin is averaged over all ten independent realisations and the error bars give the rms scatter.\label{Tinsp}}
\end{figure*}
\begin{figure*}
\centering
\includegraphics[width=\textwidth]{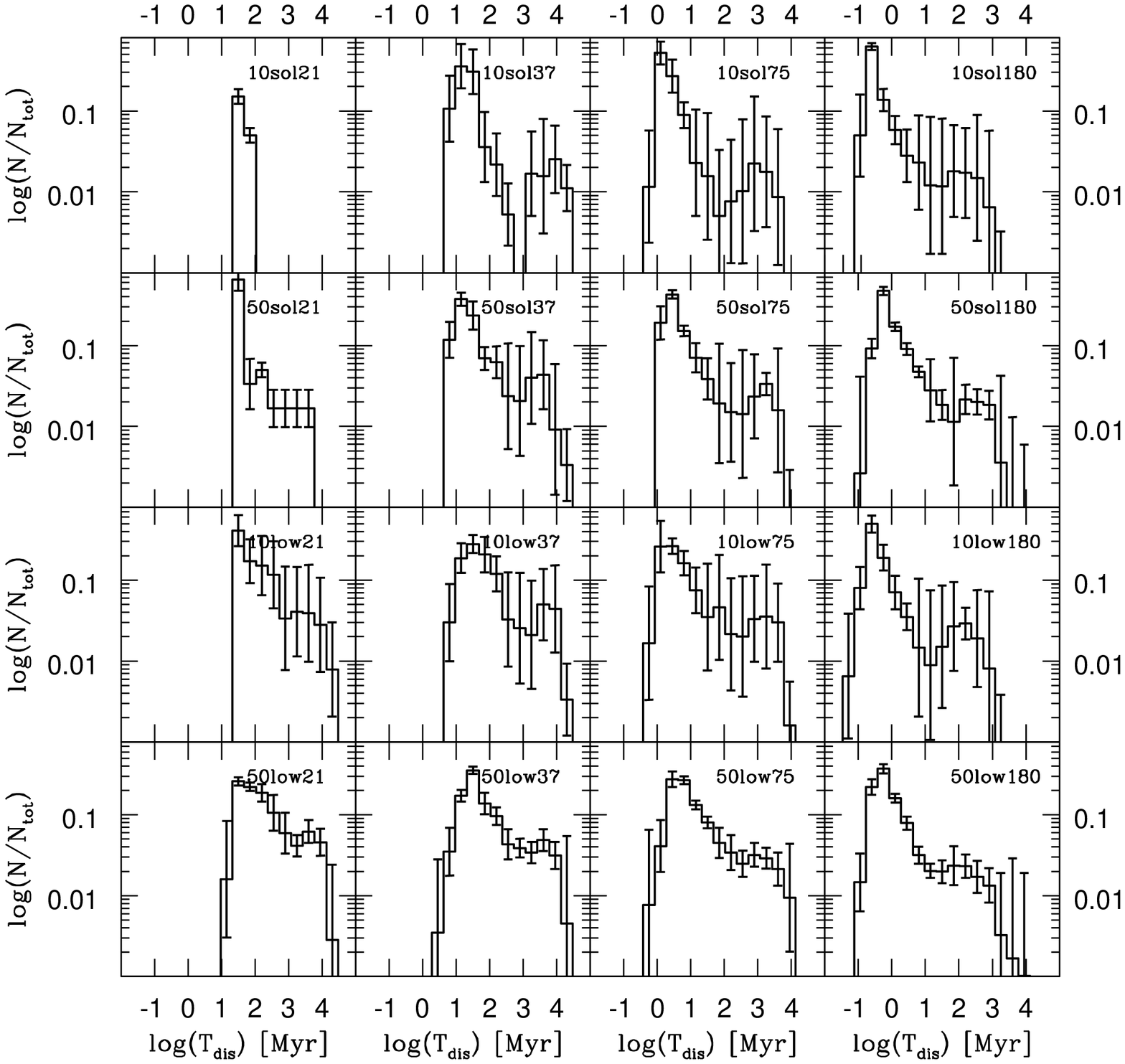}
\caption[Dynamical Disruption Timescale]{Cumulative distribution of BH-BH binary dynamical disruption timescale up to $1 T_{H}$ binned uniformly in log space.  Each bin is averaged over all ten independent realisations and the error bars give the rms scatter.\label{Lifetimes}}
\end{figure*}

In Figure~\ref{periods} we present the cumulative period ($P$) distribution of all BH-BH binaries up to $1 T_{H}$.  There is a large spread in $P$, ranging
from days to $\sim 10^{5}$ yrs.  The period distribution does not depend strongly on the cluster parameters.  The entire distribution is shifted slightly towards short periods in the more concentrated simulations.  This is partly due to the slightly higher velocity dispersion in these clusters and the consequently higher value of $k_{B}T$ in physical units.  Thus in concentrated clusters binaries must have higher binding energies and consequently shorter periods in order to be above the hard-soft boundary (such as it is in multi-mass systems).  It is also partly a product of the larger number of hardening interactions due to the higher interaction rate.  Despite the overall shift, the clusters with $r_{t}/r_{h} = 180$ show a peak in at the longer end of the distribution.  This simply reflects the peak at low binding energy seen in Figure~\ref{Ebind}.  Overall, however, the period distributions are similar and span approximately the same range for all models.

Although there are some binaries with periods less than a year present in most simulations, most binaries do not have sufficiently short periods to produce
gravitational wave mergers within $1 T_{H}$.  According to \cite{Peters64} the rate of semi-major axis decay, $\dot{a}$, for a binary emitting gravitational
waves in the orbit-averaged regime is:
\begin{equation}
  \label{PandMadot}
  \langle \dot{a} \rangle = - \frac{64}{5} \frac{G^{3}m_{1}m_{2} \left( m_{1}+m_{2} \right)}{c^{5}a^{3} \left( 1 - e^{2} \right)^{7/2}} \left( 1 + \frac  {73}{74}e^{2} + \frac{37}{96}e^{4} \right)\
\end{equation}
where $m_{1}$ is the mass of the primary, $m_{2}$ is the  mass of the secondary, $e$ is the eccentricity, and $a$ is the semi-major axis.  We can calculate an approximate inspiral time, $t_{\rm in}$, for a binary by taking:
\begin{equation}
  \label{tinsp}
  t_{\rm in} \approx \frac{a_{in}}{\dot{a}}
\end{equation}
where $a_{in}$ is the initial semi-major axis of a binary.  For a circular binary with $m_{1} = m_{2} = 10 M_{\odot}$ and an initial period of $P_{in} =
1$ day, $t_{\rm in} \approx 1$ Gyr.  If $P_{in} = 1$ yr then $t_{\rm in} = 10^6$ Gyr.  Thus all but the shortest period binaries in our clusters will be
unable to merge within $1 T_{H}$.  Furthermore, the minimum gravitational wave frequency for which LISA is sensitive is $\approx 10^{-5}$ Hz.  For circular
binaries all power is emitted in the $n = 2$ harmonic of $\omega$ \citep{PandM63,BBandB08}.  Thus for circular binaries to be detected by LISA they must have periods of less than a day.

The presence of eccentricity in a binary can vastly improve its prospect for gravitational wave detection.  \cite{PandM63} and \cite{Peters64} have shown
that with increasing eccentricity gravitational waves are preferentially emitted at higher harmonics of the orbital frequency.  This can enhance the power emitted in gravitational waves by a factor of $10^{2}$ or more for binaries with $e > 0.8$ and reduce $t_{\rm in}$ by a similar factor.  This enhances the chance of a relativistic merger within $1T_{H}$.  Emission at higher orbital harmonics produces gravitational waves with higher frequencies that would be produced by identical binaries with circular orbits.  Thus eccentricity can bring long-period binaries into the LISA band.  In Figure~\ref{PvsEcc} we show the eccentricity of our BH-BH binaries as a function of period.  It must be noted that these eccentricities are produced randomly upon creation of a binary, do not experience self-consistent dynamical evolution during interactions, and are subject to circularisation in the course of binary evolution in BSE.  Interactions tend to increase the eccentricity of binaries and thus the eccentricities produced by our simulations should be taken as lower limits. Even with this caveat, Figure~\ref{PvsEcc} shows that there are a wide range of eccentricities for any given period.

Since some of our binaries are eccentric and since this eccentricity can significantly reduce the inspiral timescale of the binary, we use Equation~\ref{tinsp} to estimate the inspiral timescale of all BH-BH binaries in our simulations.  The result is given in Figure~\ref{Tinsp} the trends of which simply reflect the period distribution in Figure~\ref{periods}.  It is apparent that even with eccentricity included there are very few binaries with an inspiral timescale shorter than $1T_{H}$.  For the few BH-BH binaries with $t_{\rm insp} < 10^{5}$ Myr the dynamics play a destructive role.  Figure~\ref{Lifetimes} shows the timescale for dynamical disruption or ejection of BH-BH binaries, $t_{\rm dis}$.  It is apparent that the average $t_{\rm dis}$ for BH-BH binaries is very short, between $1-100$ Myr in most cases, and is shorter in the more concentrated clusters due to the higher interaction rate.  There is little between the distributions in Figure~\ref{Tinsp} and Figure~\ref{Lifetimes} and $t_{\rm in} > t_{\rm dis}$ in almost all cases.  Indeed in none of our simulations do we find any BH-BH binary mergers.  Some of the eccentric, short-period binaries in our simulations may, however, have a chance of appearing in the LISA band and we will turn to this possibility in \S~\ref{sec:LISA}.

\begin{figure}
\centering
\includegraphics[width=0.5\textwidth]{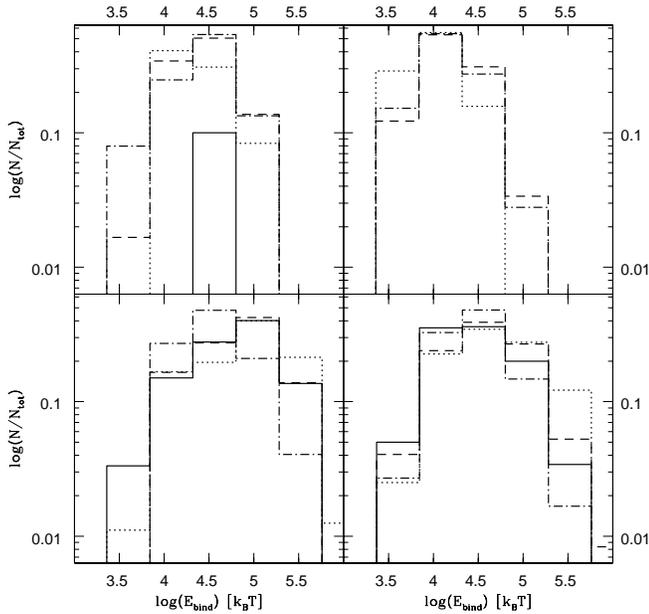}
\caption[Binding Energy of BH-BH Escapers]{The binding energy of BH-BH binary escapers binned uniformly in log space.  The top row shows the $Z = 0.02$ simulations and the bottom the $Z = 0.001$ ones.  In the right column are simulations with $f_{b} = 0.1$ and in the left simulations with $f_{b} = 0.5$ column.  Concentrations are $r_{t}/r_{h} = 21$ (solid), $r_{t}/r_{h} = 37$ (dotted), $r_{t}/r_{h} = 75$ (dashed) and $r_{t}/r_{h} = 180$ (dot-dashed).  Each bin is averaged over all ten independent realisations.\label{escape}}
\end{figure}

Although we find no mergers within the clusters, the simulations eject hard binaries.  Figure~\ref{escape} shows the distribution of binding energies for
escaping BH-BH binaries.  $E_{\rm bind}$ is, on average, much higher for the escapers than for the system as a whole.  This is because the most tightly
bound binaries tend to receive the highest recoil velocities in few-body encounters and are thus the most likely to be ejected from the system.  Therefore the most promising merger candidates are the least likely to remain in the cluster.  The distribution in Figure~\ref{escape} compares favourably with the results from \cite{Oleary06} (their Figure 6) which were derived based on explicit few-body integration.  This agreement increases our confidence that both the few-body encounters and the dense core dynamics are being re-produced successfully in our code.

\section{Gravitational Wave Detection}
\label{GWdetect}

In this section we consider the prospects for gravitational wave detection in both the ground- and space-based frequency bands.

\subsection{Ground-Based Sources}
\label{sec:LIGO}

Due to the long inspiral times and short disruption times shown in figures~\ref{Tinsp} and~\ref{Lifetimes} we find no compact mergers in any of our simulations.  Thus the BH-BH binary destruction rate within the cluster dominates the creation rate and all BH-BH binaries are either ejected or disrupted before they have a chance to inspiral and merge.  We therefore predict that globular clusters as objects will not contribute significantly to the ground-based gravitational wave detector signal.  Many of the escapers produced by GC dynamics should, however, merge in the galactic field within a Hubble time and thus may contribute to the detection rate.  We save analysis of these sources for a follow-up paper and concentrate instead on the very interesting LISA sources that we find within our simulations.

\subsection{LISA Sources}
\label{sec:LISA}

In order to study potential LISA sources from these simulations, we consider only those binaries present in the cluster simulations at ages between 10 and 14 Gyr.  This is covers the age distribution of GCs in the Milky Way and stellar mass inspirals will not be strong enough gravitational wave sources to be observed beyond the galaxy.  Advanced LIGO will be able to detect BH-BH inspirals out to moderate redshifts and for these predictions we will have to take the earlier stages of cluster evolution and the possibility of younger GCs in other galaxies into account.  Of these Milky Way binaries, we further restrict them to have a combination of orbital periods and eccentricities such that the harmonic with peak power lies within the LISA sensitivity band. We estimate the frequency of this harmonic by approximating the orbit at periastron ($a(1-e)$) with a circular orbit of radius $r=a(1-e)$. The frequency of this circular orbit is then proportional to $(1-e)^{-1.5}$. However, since the orbital speed of the eccentric binary at periapsis is higher than that of a circular binary, the dominant frequency of the gravitational wave burst will be somewhat higher than the circular frequency. For a parabolic encounter, this speed difference is $\sqrt{2}$, and so the harmonic with peak power is estimated to be
\begin{equation}
  n_{\rm max} \simeq \sqrt{\frac{2}{\left(1-e\right)^3}}.
\end{equation}
With this age and frequency/eccentricity restriction, we find that 33 simulations have potential LISA sources. All but one of these have only one potential LISA source at any time during the 10 to 14 Gyr under consideration. The one exception (a realization of the 50low37 model) has two potential LISA sources during the age span of 10.6 to 11.6 Gyr. Six of these simulations had binaries with $e < 0.9$, with one of these having $e<0.7$.  Although there are 34 potential LISA sources in these simulations,  we need to determine if these BH-BH binaries will have sufficient strength to be detected by LISA. This is done by calculating the signal-to-noise ratio using:
\begin{equation}
  \label{snrdef}
  \rho^2 = 4\int_0^\infty{\frac{\left|\tilde{h}(f)\right|^2}{S_n(f)}df}
\end{equation}
where $\tilde{h}(f)$ is the Fourier transform of the response of LISA to the gravitational wave and $S_n(f)$ is the power spectral density of the expected noise in LISA. We include both instrument noise and an estimate of the Galactic white dwarf binary foreground from \cite{Ruiter08}.

We determine the response, $h(t)$ by placing ten realisations of each binary in Galactic globular clusters that are within 5 kpc of the Earth. The properties of these clusters are obtained from \cite{Harris96}, and are shown in Table~\ref{globtable}. Each realisation is given a sky location within the globular cluster and then assigned random orientations and initial orbital phases. The barycentered waveform is determined using the harmonic expansion of  \cite{Pierro01} carried out to the $n \sim 1300$ harmonic. The response of LISA is calculated using the long wavelength approximation as described in \cite{Cutler98} for one year of observation.

\begin{table}
\centering
\caption{Celestial coordinates and distances for the 16 globular clusters
  within 5 kpc of Earth\label{globtable}}
\scriptsize{
\begin{tabular}[c]{l r r r r r r r r }
\hline
Name& \multicolumn{3}{c}{RA} & \multicolumn{3}{c}{dec} & dist  \\
{} & h & m & s & $^\circ$ & ' & " & [kpc] \\
\hline
NGC104  & 00 & 24 & 05.2 & -72 & 04 & 51 & 4.5 \\
E3      & 09 & 20 & 59.3 & -77 & 16 & 57 & 4.3 \\
NGC3201 & 10 & 17 & 36.8 & -46 & 24 & 40 & 5.0 \\
NGC6121 & 16 & 23 & 35.5 & -26 & 31 & 31 & 2.2 \\
NGC6218 & 16 & 47 & 14.5 & -01 & 56 & 52 & 4.9 \\
NGC6254	& 16 & 57 & 08.9 & -04 & 05 & 58 & 4.4 \\
NGC6366 & 17 & 27 & 44.3 & -05 & 04 & 36 & 3.6 \\
NGC6397 & 17 & 40 & 41.3 & -53 & 40 & 25 & 2.3 \\
NGC6540 & 18 & 06 & 08.6 & -27 & 45 & 55 & 3.7 \\
NGC6544 & 18 & 07 & 20.6 & -24 & 59 & 51 & 2.7 \\
2MSGC01 & 18 & 08 & 21.8 & -19 & 49 & 47 & 3.6 \\
2MSGC02 & 18 & 09 & 36.5 & -20 & 46 & 44 & 4.0 \\
Ter12   & 18 & 12 & 15.8 & -22 & 44 & 31 & 4.8 \\
NGC6656 & 18 & 36 & 24.2 & -23 & 54 & 12 & 3.2 \\
NGC6752 & 19 & 10 & 52.0 & -59 & 59 & 05 & 4.0 \\
NGC6838 & 19 & 53 & 46.1 &  18 & 46 & 42 & 4.0 \\
\hline
\end{tabular}
}
\end{table}

\begin{figure}
\centering
\includegraphics[width=0.5\textwidth]{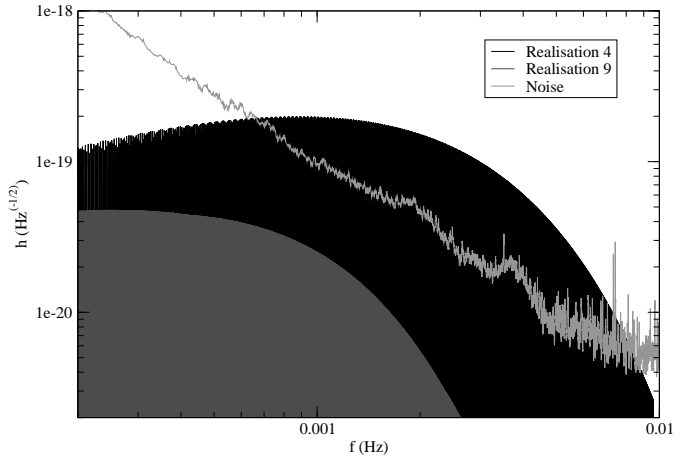}
\caption{Gravitational wave spectra of two BHBH binaries from simulation 50low37 compared with combined instrumental and Galactic white dwarf binary confusion noise. The weak signal is from realisation 9 and the strong signal is from realisation 4. Here, we have calculated the waveform out to the $n \sim 5000$ harmonic to show the full spectrum of the binaries. For the calculation of the signal-to-noise ratio, both signals are cut-off at a frequencies around 2 mHz because we only use harmonics below $n=1300$. Note that the true signal-to-noise ratio is likely higher than we have calculated.  \label{lisaspect}}
\end{figure}

Setting a detection threshold of $\rho \ge 7$ in a single interferometer (which corresponds to a combined $\rho \ge 10$ in two channels of the LISA data stream) we find potentially detectable binaries in three realisations of simulation 50low37 and  one realisation of simulation 50sol75. The 
properties of these binaries are shown in Table~\ref{LISAsources}. The notable features of these binaries are that they are highly eccentric.  The long-period binary from realisation nine of simulation 50low37 is only visible from a few orientations in the nearby globular cluster NGC6121, while the binaries in the other realisations of simulations 50low37 and 50sol75 are visible in all of the globular clusters chosen. The spectra of the binaries in realisations nine and four for simulation 50low37  from NGC 6121 are shown in figure~\ref{lisaspect} along with an estimate of the combined instrument and Galactic white dwarf binary confusion noise. It is also interesting to note that there is a preference for high primordial binary fractions, but otherwise there is no clear dependence on cluster parameters for these sources. It must again be noted that these eccentricities are not self-consistently produced by few-body calculations and must be interpreted with caution.  Furthermore, the simulations indicate at most two binaries per globular cluster with the potential for detection by LISA and the bulk of the simulations result in no detectable systems. Consequently it is difficult to make any firm predictions about the likelihood of detection of BH-BH systems in the Galactic globular cluster system with LISA. Nonetheless, if we assume that high binary-fractions are common in globular clusters and that the initial concentrations are $r_t/r_h = 37$, then roughly 30\% of nearby globular clusters may house a detectable binary.

\begin{table}
\centering
\caption{Properties of the detectable LISA sources from these
  simulations. If the orbital period decreases during the observation time, a range is given.\label{LISAsources}}
\scriptsize{
\begin{tabular}[c]{l r r r r r}
\hline
Simulation & Age & $M_1$ & $M_2$ & $P_{\rm orb}$ & $e$ \\
{} & [Gyr] & [${\rm M}_\odot$] & [${\rm M}_\odot$] & [$\times 10^3$ s] & {} \\
\hline
50low37.4  & $< 10.2$ & 20.5 & 25.9 &  920--490 & 0.988 \\
50low37.9 & $> 12.5$ & 11.1 & 16.8 & 1190 & 0.986 \\
50low37.10 & $ >11.6$ & 28.4 & 14.3 & 215 & 0.933\\
\hline
50sol75 & $<10.1$ & 10.8 & 8.5 & 479--257 & 0.998 \\
\hline
\end{tabular}
}
\end{table}

\section{Discussion}
\label{discussion}

Although we produce no compact binary mergers within our simulations, we can still compare our dynamics to \cite{Oleary06} and \cite{Sadowski08}.  As described in \S~\ref{intro}, \cite{Oleary06} assume that the BHs are sufficiently mass-segregated that they form a decoupled subsystem at the centre of the cluster and interact only with themselves.  By contrast \cite{Sadowski08} assume that the BHs always remain in equilibrium with the rest of the cluster and, since they do not become centrally concentrated, remain at much lower density than in the mass-segregated case.  Therefore the BH-BH binaries in
the \cite{Oleary06} simulations frequently interact with each other whereas the BH-BH binaries in \cite{Sadowski08} do not.  This means that the BH-BH binary formation rate reported in \cite{Oleary06} is much higher than that reported in \cite{Sadowski08}.  However, since the BH-BH binaries are the only objects in the system massive enough to disrupt and eject other BH-BH binaries, the BH-BH binaries in the \cite{Oleary06} simulations are much more likely to be disrupted than the binaries in the \cite{Sadowski08} simulations.  In practice the lower disruption rate wins and \cite{Sadowski08} produces a larger and more constant rate of BH-BH mergers than does \cite{Oleary06}.  Therefore \cite{Oleary06} provides a lower limit on gravitational wave detection rates and \cite{Sadowski08} provides an upper limit.

Our simulations, including full global dynamics, suggest that the \cite{Oleary06} approximation is more accurate.  Figure~\ref{radii} shows that the BHs in our simulations mass-segregate swiftly and that BH-BH binaries form in the centre of the cluster.  The short dynamical disruption timescales in Figure~\ref{Lifetimes} imply that the BH-BH binaries interact strongly with each other and are swiftly destroyed.  The binding energy distribution of escapers in  Figure~\ref{escape} shows that hard binaries are preferentially ejected from the cluster, suggesting a dynamical ejection scenario.  Finally the time-dependant number of BH-BH binaries per Gyr reported in Figure~\ref{BHBHrates}, particularly for the dense clusters, is more consistent with the time-dependent merger rates of \cite{Oleary06} than the constant merger rate found in \cite{Sadowski08}.

A major difference between our results and those of both \cite{Oleary06} and \cite{Sadowski08} is that we find no BH-BH mergers within our simulated clusters.  \cite{Oleary06} find at least 30\% of their mergers occur within the clusters and \cite{Sadowski08} find 90\%.  The difference between \cite{Oleary06} and \cite{Sadowski08} is due to the larger number of interactions and hence the larger number of ejections in the \cite{Oleary06} simulations.  Our lack of mergers within the clusters is a result our high interaction rates but is also affected by the more approximate treatment of few-body interactions in our Monte Carlo code.  In particular our binary-binary interaction prescriptions follow those of \cite{Stod86} and in these prescriptions 88\% of all interactions result in the disruption of the softer binary.  This works well for normal binary-binary interactions but may over-predict the disruption rate in interactions between two hard BH-BH binaries.  In reality more of these binaries should probably survive and be hardened by the interaction.  The prescriptions also gives $0.516(E_{b1}+E_{b2})$, where $E_{b1,2}$ are the binding energies of the two binaries, as increased binding energy to the hard binary and distributes the same amount as kinetic energy between the centres of mass after the interaction.  In general this produces the correct hardening of the surviving binary (compare the energy distribution in Figure~\ref{escape} to Figure 6 in \cite{Oleary06}) but could well produce recoil velocities and hence escape rates that are systematically to high.  Furthermore these prescriptions allow neither mergers during the interaction nor long-lived hierarchical triples.  The combination of all these effects means that the number of mergers within our clusters is almost certainly too low.

Our simulations represent the first quantitative study of stellar-mass BH-BH binaries as LISA sources within star clusters.  We can compare our results to the number of stellar-mass sources predicted in the galactic field population by \cite{BBandB08} and quoted in \S~\ref{intro}.  When normalised to the total mass in their simulations, \cite{BBandB08} find, in the case where mergers on the Hertzsprung gap are not allowed, $2 \times 10^{-5}$ resolvable LISA sources in the Milky Way per $10^{5} M_{\odot}$, $1 \times 10^{-5}$ of which per $10^{5} M_{\odot}$ are BH-BH.  In the case where mergers during the Hertzsprung gap are allowed they find $4 \times 10^{-6}$ per $10^{5} M_{\odot}$, of which none are BH-BH. Assuming all four of the sources in our simulations are resolved at the current time this yields $6 \times 10^{-3}$ detections per $10^{5} M_{\odot}$ from our simulations, all BH-BH.  This rate depends on the parameters of the individual binaries but has no clear dependence on the cluster parameters.  All sources are highly eccentric, indicating that eccentric stellar mass sources may be present in the LISA band.  Many of our escaping binaries also have periods of a day or less and will almost certainly appear in the LISA band at some point.  This does, however, place a limitation on the interpretation of the results of \cite{BBandB08}.  In their model the presence of stellar-mass BH-BH binaries in the LISA band would indicate that mergers  in the Hertzsprung gap are rare whereas our results show that even if mergers on the Hertzsprung gap are common, BH-BH binaries could still exist in the LISA band due to dynamical processes in star clusters.  However, the BH-BH binaries generated through dynamical processes will be associated with individual globular clusters if they are retained, or in the halo if they have been ejected. This is a distinct population from the disk population of binaries found in \citet{BBandB08}.

Finally we note in passing that although we have analysed our simulations in terms of compact binaries they are in no way limited to such studies.  Stellar evolution is calculated for all stars in the cluster and thus our simulations represent a complete star cluster populations synthesis study based on the stellar evolution tracks of \cite{HPandT00} and \cite{HTandP02}.  We hope to make the full results of our simulations publicly available in the near future and encourage anyone who is interested in such data to contact the authors.

\section{Conclusions}
\label{conclude}

We have studied the dynamics of the BH population in star clusters with a self-consistent Monte Carlo treatment of the global dynamics and full stellar and binary evolution.  We confirm the predictions of \cite{SigPhin93} that BH-BH, but not NS-NS or NS-BH, binaries are produced efficiently in star clusters by dynamical interactions.  We find that the BHs are mass-segregated and interact strongly with each other, confirming the more approximate models of \cite{Oleary06}.  We find no BH-BH mergers within the clusters we simulate but many hard BH-BH escapers that will merge in the galactic field within a Hubble time.  Detection rates for both ground- and space-based detectors will be calculated and presented in a future paper.  We find that our simulations produce potential LISA sources that, while rare, will be highly eccentric and may still represent a significant enhancement to the galactic field population.  We will certainly produce more detections when the escapers are included.  We conclude that star clusters produce BH-BH binaries efficiently and must be taken into account when considering detection rates for both ground- and space-based detectors.

\section*{Acknowledgements}

J.M.B.D. would like to thank the International Max-Planck Research School for Astronomy and Cosmic Physics at the University of Heidelberg (IMPRS-HD) for providing funding for his Ph.D.  The simulations have been carried out at the High Performance Computing Center Stuttgart (HLRS) using the resources of Baden-W\"u{}rttemberg grid (bwgrid) through the German Astrogrid-D and D-Grid projects.  M.J.B acknowledges the support of NASA Grant NNX08AB74G and the Center for Gravitational Wave Astronomy, supported by NSF award \#{}0734800.  M.G. was supported by Polish Ministry of Science and Higher Education through the grant 92/N.ASTROSIM/2008/0 and N N203 380036.  R.S. thanks the Deutsches Zentrum f\"u{}r Luft- und Raumfahrt (DLR) for support within the LISA Germany project.

\label{lastpage}

\end{document}